\documentclass[a4paper,notitlepage,12pt,thmsb]{article}
\usepackage{amssymb}

\usepackage{epsfig}
\usepackage{amsmath}
\usepackage{graphicx}

\begin{document}

\title{\textbf{Origin of causal set structure in the quantum universe}}
\author{Jon Eakins and George Jaroszkiewicz \\
\\
School of Mathematical Sciences, University of Nottingham,\\
University Park, Nottingham NG7 2RD, UK}
\maketitle

\begin{abstract}
We discuss the origin of causal set structure and the emergence of classical
space and time in the universe. Given that the universe is a closed
self-referential quantum automaton with a quantum register consisting of a
vast number of elementary quantum subregisters, we find two distinct but
intimately related causal sets. One of these is associated with the
factorization and entanglement properties of states of the universe and
encodes phenomena such as quantum correlations and violations of Bell-type
inqualities. The concepts of separations and entanglements of states are
used to show how state reduction dynamics generates the familial
relationships which gives this causal set structure. The other causal set
structure is generated by the factorization properties of the observables
(the Hermitian operators) over the quantum register. The concept of skeleton
sets of operators is used to show how the factorization properties of these
operators could generate the classical causal set structures associated with
Einstein locality.

\vspace{0.25in}

\noindent PACS number(s): 03.65.Ca, 03.65.Ud, 04.20.Gz, 98.80.Qc
\end{abstract}

\newpage

\begin{center}
{\large {\textbf{I. INTRODUCTION}} }
\end{center}

Whilst attempting to account for the existence of space, time and matter in
the universe, physicists often adopt one of two opposing viewpoints. These
may be labelled \textit{bottom-up} and \textit{top-down} respectively,
reflecting the basic difference between reductionist and holistic physics.
Such a difference is to be expected given a universe running on quantum
principles, because quantum mechanics (\textit{QM}) is simultaneously
reductionist and holistic.

Many bottom-up approaches proceed from the assertion that at its most basic
level, the universe can be represented by a vast collection of discrete
events embedded in some sort of mathematical space, such as a manifold. By
avoiding any assumption of a pre-existing manifold, the \textit{pre-geometric%
} approach goes further and asserts that conventional classical time and
space and the classical reality that we appear to experience all emerge on
macroscopic scales due to the complex connections between these fundamental
microscopic entities. This approach was pioneered by Wheeler \cite
{WHEELER-80} and has received attention more recently by Stuckey \cite
{STUCKEY-99}.

On the other hand, most top-down approaches to quantum cosmology consider
the universe as a single quantum system with a unique state evolving
according to a given set of laws or conditions. From this point of view, our
classical picture of the universe is no more than a good working
approximation whenever a detailed quantum mechanical description can be
neglected.

In this paper we apply a theoretical framework or paradigm which attempts to
explain and reconcile these viewpoints within a consistent set of postulates
based on the principles of standard quantum mechanics, stretched to cover
the unique case when the system under observation is the universe and not
just a subsystem of it. In this paradigm, reviewed briefly in the Appendix,
discrete structures occur naturally, being generated by the factorization
and entanglement properties of states and observables.

Related to our approach is the bottom-up approach to cosmology known as the
causal set hypothesis \cite
{SORKIN+al-87,BRIGHTWELL-91,SORKIN-99,SORKIN+RIDOUT-00,BRIGHTWELL-02}. This
assumes that spacetime is discrete at the most fundamental level. In those
models it is postulated that classical, discrete events are generated at
random, though neither the nature of these events nor the mechanism
generating them is explained or discussed in detail. In this paper we
explain how the particular mathematical properties and dynamical principles
of quantum theory can generate such a structure. A particularly important
feature of our approach is the occurrence of two distinct causal set
structures, in contrast to the one normally postulated. One of these causal
sets arises from the entanglement and separation properties of states whilst
the other arises from the separation and entanglement properties of the
operators. We believe that these two causal sets correspond to the two sorts
of information transmission observed in physics, that is, non-local quantum
correlations which do not respect Einstein locality and classical
information transmissions which do respect it.

Although it seems natural to generate causal sets by the discretization of a
pseudo-Riemannian spacetime manifold of fixed dimension, as is done in
lattice gauge theories and the Regge discretization approach \cite{REGGE-61}
to general relativity ($GR$), causal set events need not in principle be
regarded as embedded in some background space of fixed dimension $d$.
Instead, conventional (i.e. physical) spacetime is expected to emerge in
some appropriate limit as a consequence of the causal set relations between
the discrete events. It is expected that in the correct continuum limit,
metric structure should emerge \cite{BRIGHTWELL-91}. Additionally, it has
been suggested \cite{SORKIN+al-87}\ that the dimension of this emergent
spacetime might be a scale dependent quantity, making the model potentially
compatible with general relativity in four spacetime dimensions, string and
m-brane cosmology and higher dimensional Kaluza-Klein theories. The
attractive feature here is the idea of some universal ``pregeometric''
structure capable of accounting for all known emergent features and from
which most, if not all, currently popular models of the universe would
emerge as reasonable approximations in the right contexts.

\ Important contemporary issues in quantum mechanics which impact directly
on quantum cosmology and hence on our work are the status of state reduction
\cite{PLENIO-98} and the physical meaning of time. For a number of reasons,
state reduction has often been regarded as an unattractive modification of
more elegant quantum principles based on Schr\"{o}dinger evolution. In the
many worlds and decoherence interpretations of quantum mechanics for
instance, quantum jumps are regarded in principle as non-existent, at best
being useful approximations to a underlying continuous quantum evolution.
The physical meaning of time is intimately bound up with this issue, because
in the physics laboratory and in the sphere of more general human activity,
physical time is marked by irreversible processes of information acquisition
and loss, the origin of which is normally attributed to state reduction.

Up to relatively recently, it was reasonable to follow Schr\"{o}dinger in
his view that quantum mechanics is a theory dealing with ensembles, with the
consequence that wavefunctions should be regarded as no more than an
encoding of statistical information. Certainly, there has been in recent
years a view amongst many leading theorists that quantum states should not
be regarded as real attributes of systems.

This impacts on the interpretation of time. Cutting a long story short,
there are two distinct views of time, which we will refer to as the manifold
time and process time perspectives respectively. The former considers time
as part of a four dimensional continuum for which relativistic principles
hold. Physical theories have to be Lorentz covariant in their predictions of
local expectation values and there is no inherent difference between past
and future on the most superficial levels of the theory. This leads to the
block universe picture of a universe in which structural (i.e., geometrical)
properties hold predominantly.

The process view of time relates to dynamical processes of change. According
to this view, only the moment of the present has physical meaning and the
future can only be discussed in terms of potential. It is well understood
that this perspective does not square with relativistic principles, because
in relativity there is no absolute simultaneity. However, both perspectives
are consistent with quantum principles. In the manifold perspective, we can
use quantum mechanics to calculate expectation values for past, present and
future, if we ignore the thorny issue of exactly what goes on when
information is extracted from physical systems (the measurement problem).
From the process time perspective, quantum mechanics gives probabilities for
individual future outcomes of quantum tests.

Consistent with the notion that single outcomes can be physically relevant
in quantum mechanics are the recent developments in quantum physics
concerning single ion traps $\cite{PLENIO-98}$. There, a single subsystem is
subject to external probing in such a way that the best description of what
is going on is in terms of quantum jumps. It seems no longer possible to
maintain Schr\"{o}dinger's view that we can never experiment on single
particles but only on ensembles.

These experimental developments support the case that, contrary to what
decoherence asserts, quantum state reduction processes are meaningful.
Indeed, state reduction is as equally important in quantum computation as
the unitary evolution characterizing the relationship between initial and
final states and the decohering influence of the environment.

\

The present paper incorporates the notion of an underlying qubit pregeometry
with state reduction into a specific paradigm, which for convenience we
refer to as the stages paradigm. We believe that the mathematical structures
occurring naturally in quantum registers have the potential to answer a
number of questions important to quantum physics, such as the relationship
of the observer to the system under observation. Many directions still have
to be explored. In this paper we will lay down general ideas, explaining how
a causal set structure can arise naturally within the stages paradigm, and
how much of the Hasse diagrams generated in causet theory may be recreated
as the state of the universe $\Psi _{n}$ factorizes and re-factorizes over
successive jumps. We believe, however, that the classically motivated line
of thinking in \cite{SORKIN-99} is too general, because whilst it may seem
mathematically possible to produce any configuration of elements in a
causet, not all types of relationship specific to classical Hasse diagrams
are permissible in our approach to quantum physics unless some sort of
external `information store' is available. Conversely, evolution of states
in the stages paradigm will be shown to reproduce only those parts of the
Hasse diagrams that are allowed by quantum mechanics and are hence
physically meaningful.

\begin{center}
\textbf{A. Plan}
\end{center}

In $\S 2$ we review classical causal set theory, followed by a discussion of
some points relevant to quantum cosmology in $\S 3$. In $\S 4$ we introduce
the notions of separations and entanglements, which describe the possible
partitions of a direct product Hilbert space into subsets of relevance to
physics. These concepts provide a natural basis for causal set structure
once dynamics is introduced. This is discussed in $\S 5$, where we show how
state reduction concepts inherent to the stages paradigm lead to a natural
definition of the concepts of families, parents and siblings used in causal
set theory. In $\S 6$ we discuss how two sorts of causal set structure arise
in our paradigm, one associated with states and the other with observables,
and explain how these are related to non-local quantum correlations and
Einstein locality respectively. In $\S 7$, we introduce the notion of \emph{%
skeleton set}, used to construct observables, and discuss the concepts of
separability and entanglement for operators. In $\S 8$ we discuss the
fundamental role of eigenvalues in the theory, in $\S 9$ we discuss some
physically motivated examples, followed by our summary and conclusions in $%
\S 10$. In the Appendix we review the stages paradigm, which is central to
our work.

\

\begin{center}
{\large {\textbf{II. CAUSAL SETS}} }
\end{center}

A number of authors \cite
{SORKIN+al-87,BRIGHTWELL-91,SORKIN-99,MARKOPOULOU-00,REQUARDT-99} have
discussed the idea that spacetime could be discussed in terms of causal
sets. In the causal set paradigm, the universe is envisaged as a set $%
\mathcal{C}\equiv \{x,y,\ldots \}$ of objects (or \emph{events}) which may
have a particular binary relationship amongst themselves denoted by the
symbol\ $\prec ,$ which\ may be taken to be a mathematical representation of
a temporal ordering. For any two different elements $x,y$, if neither of the
relations $x\prec y$ nor $y\prec x$ holds then $x$ and $y\,$ are said to be
\emph{relatively spacelike}$,$ \emph{causally independent} or \emph{%
incomparable }\cite{HOWSON:72}. The objects in $\mathcal{C}$ are usually
assumed to be the ultimate description of spacetime, which in the causal set
hypothesis is often postulated to be discrete \cite{SORKIN-99}. Minkowski
spacetime is an example of a causal set with a continuum of elements \cite
{BRIGHTWELL-91}, with the possibility of extending the relationship $\prec $
to include the concept of \emph{null} or \emph{lightlike} relationships.

The causal set paradigm supposes that for given elements $x,y,z$ of the
causal set $\mathcal{C}$, the following relations hold:
\begin{align}
\forall\,x,y,z & \in\mathcal{C},\;\;\;\;x\prec y\;\mathrm{and\;}y\prec
z\Rightarrow x\prec z\;\;\;\mathrm{(transitivity)}  \notag \\
\forall x,y & \in\mathcal{C},\;\;\;\;\;x\prec y\Rightarrow y\nprec
x\;\;\;\;\;\;\;\;\;\;\;\;\mathrm{(asymmetry)} \\
\forall x & \in\mathcal{C},\;\;\;\;\;x\nprec
x.\;\;\;\;\;\;\;\;\;\;\;\;\;\;\;\;\;\;\;\;\;\;\mathrm{(irreflexivity)}
\notag
\end{align}
A causal set may be represented by a Hasse diagram \cite{HOWSON:72}. In a
Hasse diagram, the events are shown as spots and the relations as solid
lines or links between the events, with emergent time running from bottom to
top.

One method of generating a causal set is via a process of `sequential
growth' \cite{SORKIN-99}. At each step of the growth process a new element
is created at random, and the causal set is developed by considering the
relations between this new event and those already in existence.
Specifically, the new event $y$ may either be related to another event $x$
as $x\prec y$, or else $x$ and $y$ are said to be unrelated. Thus the
ordering of the events in the causal set is as defined by the symbol $\prec,
$ and it is by a succession of these orderings, i.e. the growth of the
causet, that constitutes the passage of time. The relation $x\prec y$ is
hence interpreted as the statement: ``$y$ is to the future of $x$''.
Further, the set of causal sets that may be constructed from a given number
of events can be represented by a Hasse diagram of Hasse diagrams \cite
{SORKIN-99}.

The importance of causal set theory is that in the large scale limit of very
many events, causal sets may yield all the properties of continuous
spacetimes, such as metrics, manifold structure and even dimensionality, all
of which should be determined by the dynamics \cite{SORKIN+al-87}. For
example, it should be possible to use the causal order of the set to
determine the topology of the manifold into which the causet is embedded
\cite{BRIGHTWELL-91}. This is the converse of the usual procedure of using
the properties of the manifold and metric to determine the lightcones of the
spacetime, from which the causal order may in turn be inferred.

Distance may be introduced into the analysis of causal sets by considering
the length of paths between events \cite{SORKIN+al-87,BRIGHTWELL-91}. A
\emph{maximal chain} is a set $\left\{ a_{1},a_{2},...,a_{n}\right\} $ of
elements in a causal set $\mathcal{C}$ such that, for $1\leq i\leq n$, we
have $a_{i}\prec a_{i+1}$ and there is no other element $b$ in $\mathcal{C}$
such that $a_{i}\prec b\prec a_{i+1}.$ We may define the path length of such
a chain as $n-1$. The distance $d(x,y)$ between comparable \cite{HOWSON:72}
elements $x,y$ in $\mathcal{C}$ may then be defined as the maximum length of
path between them, i.e. the `longest route' allowed by the topology of the
causet to get from $x$\ to $y$. This implements Riemann's notion that
ultimately, distance is a counting process \cite{SORKIN+al-87}. For
incomparable elements, it should be possible to use the binary relation $%
\prec$ to provide an analogous definition of distance, in much the same way
that light signals may be used in special relativity to determine distances
between spacelike separated events.

In a similar way, ``volume'' and ``area'' in the spacetime may be defined in
terms of numbers of events within a specified distance. Likewise, it should
be possible to give estimates of dimension in terms of average lengths of
path in a given volume. An attractive feature of causal sets is the
possibility that different spatial dimensions might emerge on different
physical scales \cite{SORKIN+al-87}, whereas in conventional theory, higher
dimensions generally have to be put in by hand.

\

Our approach differs from the above in certain important respects. In the
stages paradigm, reviewed in the Appendix, spacetime \emph{per se} does not
exist and therefore cannot be regarded as being discrete or otherwise. Our
discrete sets arise naturally from the separation and entanglement
properties of quantum states and operators and are therefore not related
directly to discrete space or to the discretization of space.\

Furthermore, in the stages paradigm, various relations assumed in
``sequential growth'' must be interpreted carefully. In quantum physics,
past, present and future can never have equivalent status. At best we can
only talk about conditional probabilities, such as asking for the
probability of a possible future stage \emph{if} we assumed we were in a
given present stage. This corresponds directly to the meaning of the Born
interpretation of \ probability in \emph{QM}, where all probabilities are
conditional: \emph{if} we were to prepare $\Psi $, then the conditional
probability of subsequently detecting $\Phi $ is given by $P(\Phi |\Psi
)\equiv |\langle \Phi |\Psi \rangle |^{2}$. This does not mean that we
actually have to prepare $\Psi $.

Another problem is that, as they stand, the classical causal set relations
discussed above suggest that the various elements $x,y,z$ have an
independent existence outside of the relations themselves and that these
relations merely reflect some existing attributes. This is a ``block
universe'' perspective \cite{PRICE:97} which runs counter not only to the
process time perspective but also to the basic principles of quantum
mechanics, the implications of which lead to the uncertainty principle,
violations of Bell inequalities and the Kochen-Specker theorem \cite
{PERES:93}. All of these support Bohr's view that the quantum analogues of
classical values such as position and momentum do not exist independently of
observation.

A further criticism of classical causal sets from the point of view of
quantum theory comes from an interpretation of what the Hasse diagrams
actually represent. In some diagrams, relatively spacelike events with no
previous causal connection are permitted to be the parents of the same
event. The problem is, given two such unrelated events at a given time $n,$
it is not clear how any information from either of them could ever coincide,
that is, be brought together to be used to create any mutual descendants
unless there is some external agency organizing the flow of that
information. The whole point of causal set theory, however, is that there is
no external space in which these events are embedded, or any external
``memory'', observer or information store correlating such information, and
so it is not clear how such processes could be encoded into the dynamics. In
our approach, there are actually two dynamically interlinked causal sets,
rather than one,with different but interlinked properties stemming from the
underlying Hilbert space structure, and this solves this particular problem.

\

\begin{center}
{\large {\textbf{III. QUANTUM COSMOLOGY}} }
\end{center}

Causal sets were devised as explanations of how the universe might run
classically. The introduction of quantum mechanics into the discussion leads
to a picture of the universe as a vast quantum automaton or quantum
computation. However, not all authors agree that there exists an explicit
wavefunction for the universe. Fink and Leschke \cite{FINK+LESCHKE-00}
argued that the universe cannot be treated as a complete quantum system
because by definition, the universe cannot be part of an ensemble, nor can
it have any sort of external observer. Other authors \cite{BREUER-95} have
argued that there is no objective meaning to the notion of a quantum state
\emph{per se} other than in the context of measurement theory with
exo-physical observers.

Our counter arguments are based principally on the lack of evidence for any
identifiable ``Heisenberg cut'', or dividing line, between classical and
quantum perspectives. From the subatomic to galactic scales, when looked at
carefully, every part of the universe seems to be described by quantum
mechanics. In view of the overwhelming empirical success of quantum
mechanics on all scales, the conclusion we draw is that quantum principles
must govern everything, including the universe itself. In the particular
case of the universe, it must be regarded as a unique system for which the
conventional quantum rules concerning observers cannot be applied, because
they were only ever formulated with reference to true subsystems of the
universe. For all such subsystems, an ``exo-physical'' quantum mechanical
description is possible (where the observer looks into a subsystem ``from
the outside''), but for the unique case of the universe, an
``endo-physical'' description (where all observers are part of the system)
can be physically meaningful.

An important corollary to this line of thought is that if indeed the
universe may be represented by a quantum state, then that state can only be
a pure state. Any ``state of the universe'' cannot be a mixed state, because
there is no physical meaning to classical uncertainty in this context. The
universe cannot be part of an ensemble as far as the present is concerned,
and the only option therefore is to use a pure state description. This is a
central tenet of our approach, discussed in the Appendix.

In the long term, it will be necessary to give a detailed account of how
such pure states could be used to discuss conventional quantum physics. That
is an aspect of our work which is related to the problem of \emph{emergence}%
, i.e., an account of how the world that we think we see arises from a more
fundamental quantum basis. This is a difficult programme which we cannot
comment further on here, save to say that in our approach, we expect the
factorization properties of states and operators will give a handle on the
issue.

Given that the universe is a complete quantum system, there remains the
fundamental issue of state reduction (wave-function collapse) versus
Schr\"{o}dinger evolution, with a sharp split between those who do believe
in state reduction and those that do not. The many-worlds paradigm \cite
{DeWITT+GRAHAM:73} and its developments \cite{DEUTSCH:97,DEUTSCH-01}
explicitly rule out state reduction, as do various quantum cosmological
theories and the general framework known as decoherence. On the other hand,
the standard position stated in $QM$ texts \cite{PERES:93}, assumed here to
be the origin of the notion of time as a dynamical process \cite
{JAROSZKIEWICZ-01A}, takes at face value the role of information acquisition
as the true meaning of time. To quote Wheeler \cite{WHEELER-80}, ``\emph{The
central lesson of the quantum has been stated in the words `No elementary
phenomenon is a phenomenon until it is an observed (registered) phenomenon'''%
}.

We take the position that if the universe is described by a deterministic
Schr\"{o}dinger equation, with no probabilistic interpretation attributed to
the wave-function, then there is no way that true quantum randomness could
ever emerge. Any discussion of quantum randomness in such a paradigm would
be a pseudo-randomness based on emergent approximations, such as partial
tracing over various degrees of freedom so as to provide some sort of
justification for the appearance of mixed states in the theory. It is
acknowledged by the practitioners of many-worlds and decoherence that
accounting for the Born probability rule presents a serious problem in those
paradigms. Moreover, because all the structures in the universe are
deterministic in such paradigms, then even such approximations and their
apparent random outcomes would be predetermined. We take it as self-evident
that genuine randomness cannot occur in any system based entirely on
deterministic equations. The only option left within such systems is to
introduce ad-hoc elements such as semi-classical observers with free will.

\newpage

\begin{center}
{\large {\textbf{IV. SPLITS, PARTITIONS, SEPARATIONS AND ENTANGLEMENTS}} }
\end{center}

One of the features of quantum mechanics which distinguishes it from
classical mechanics is the occurrence of entangled states. However, as we
have stressed in \cite{JAROSZKIEWICZ-03A}, an equally important feature is
the existence of separable states, i.e., states which are direct products of
more elementary states and therefore are not entangled. Some of the factors
of such states are used by physicists to represent subsystems under
observation, whilst other factors are often used to represent the
environment and pointer states of apparatus. In our paradigm, the
possibility of various factors remaining relatively unchanged as the
universe jumps should account for the occurrence of large scale structures
having a ``trans-temporal'' identity of sorts (in a statistical sense), long
enough for classical descriptions to be applicable to them. Some of these
structures would to all intents and purposes behave as semi-classical
observers, whilst others would be identified with systems under observation.
In our paradigm, there is no inherent difference between the concepts of
observer and system, except possibly for the scales associated with them. If
the universe is describable by a tensor product Hilbert space consisting of
more than $10^{180}$ qubits, as we estimate \cite{JAROSZKIEWICZ-03A}, then
there are sufficiently many degrees of freedom to describe vast numbers of
different ``observers'' and vast numbers of different ``systems''.

In our paradigm, separations and entanglements play equally fundamental
roles and it becomes necessary to introduce a convenient notation to discuss
them, as follows.

By definition, tensor product Hilbert spaces contain both separable and
entangled elements. We shall call such a space a quantum register. We base
our discussion on a finite dimension quantum register $\mathcal{H}_{\left[
1\ldots N\right] }$ which is the tensor product
\begin{equation}
\mathcal{H}_{\left[ 1\ldots N\right] }\equiv \mathcal{H}_{1}\otimes \mathcal{%
H}_{2}\otimes \ldots \otimes \mathcal{H}_{N}  \label{111}
\end{equation}
of a vast number $N$ $\ $of factor Hilbert spaces $\mathcal{H}_{i}$, $%
1\leqslant i\leqslant N$, each known as a quantum subregister. The dimension
$d_{i}$ of the $i^{th}$ quantum subregister will be assumed to be prime.
When this dimension is two, such a subregister is known as a quantum bit, or
qubit. We restrict our attention to quantum sub-registers of prime dimension
because we shall suppose that any Hilbert space which has a dimension $d$ $%
=pq,$ where $p$ and $q$ are integers greater than one, is isomorphic to the
tensor product of two Hilbert spaces of dimensions $p$ and $q$ respectively.

In our usage of the tensor product symbol $\otimes $, the ordering is not
taken to be significant. Left-right ordering is conventionally used as a
form of labelling in differential geometry, for example, but in the context
of three or more subregisters, entanglements can occur between elements of
any of the subregisters. On account of this it is better to use subscript
labels such as in (\ref{111}) to identify specific subregisters (which are
therefore regarded as having their own identities), rather than left-right
position. For instance, if $(j_{1}j_{2}\ldots j_{N})$ is any permutation of $%
(12\ldots N)$ then we may write $(\ref{111})$ in the equally valid form
\begin{equation}
\mathcal{H}_{\left[ 1\ldots N\right] }=\mathcal{H}_{j_{1}}\otimes \mathcal{H}%
_{j_{2}}\otimes \ldots \otimes \mathcal{H}_{j_{N}}.  \label{118}
\end{equation}
There is therefore no natural ordering of quantum sub-registers in our
approach, and in the long run this is related to quantum non-locality. In
any case, any proposed ordering would have two clear problems. First, it
would suggest that any given subregister was ``further away'' from some
subregisters than others, and secondly, there is no obvious criterion for
making such an ordering anyway. On account of this lack of intrinsic
ordering, it is important to understand that the quantum subregisters are
not regarded \emph{a priori} as being embedded in any way in some
pre-existing manifold.

We shall call this the non-locality property of our tensor products. This
property holds for states as well as sub-registers. For example, if $|\Psi
\rangle \equiv |\psi \rangle _{1}\otimes |\phi \rangle _{2}$ is a separable
element of $\mathcal{H}_{\left[ 12\right] }\equiv \mathcal{H}_{1}\otimes
\mathcal{H}_{2},$ such that $|\psi \rangle _{1}\in \mathcal{H}_{1}$ and $%
|\phi \rangle _{2}\in \mathcal{H}_{2}$, then we may equally well write $%
|\Psi \rangle =|\phi \rangle _{2}\otimes |\psi \rangle _{1}.$

\begin{center}
\textbf{Splits}
\end{center}

A \emph{split} is any convenient way of grouping the $N$ subregisters in $%
\mathcal{H}_{\left[ 1\ldots N\right] }$ into two or more factors, each of
which is itself a tensor product of subregisters and therefore a vector
space. The nonlocality property of tensor products in general permits many
different splits of the same quantum register. For example, we may write the
register $\mathcal{H}_{\left[ 123\right] }$ in $5$ different ways:
\begin{eqnarray}
\mathcal{H}_{\left[ 123\right] } &=&\mathcal{H}_{1}\otimes \mathcal{H}_{%
\left[ 23\right] }=\mathcal{H}_{3}\otimes \mathcal{H}_{\left[ 12\right] }=%
\mathcal{H}_{2}\otimes \mathcal{H}_{\left[ 13\right] }  \notag \\
&=&\mathcal{H}_{1}\otimes \mathcal{H}_{2}\otimes \mathcal{H}_{3}.
\end{eqnarray}
Splits are important to our quantum causal sets because the number of
families in a given transition amplitude equals the number of factors in the
corresponding split of the total quantum register. In general, the number of
ways of splitting a register with $n$ subregisters is given by the $n^{th}$
Bell number $B_{n}$, which satisfies the recursion relations
\begin{equation}
B_{n+1}=\sum_{k=0}^{n}\tbinom{n}{k}B_{k},\;\;\;B_{0}=1,
\end{equation}
the solution of which is given explicitly by Dobinski's formula
\begin{equation}
B_{n}=\frac{1}{e}\sum_{k=0}^{\infty }\frac{k^{n}}{k!}.
\end{equation}
This sequence grows rapidly with $n$, which indicates that quantum causal
sets can have very complex structure.

\begin{center}
\textbf{Partitions}
\end{center}

Now $\mathcal{H}_{\left[ 1\ldots N\right] }$ is a vector space of dimension $%
d=d_{1}d_{2}\ldots d_{N}$ which contains both entangled and separable
states, but this classification of all states in $\mathcal{H}$ into
separable or entangled sets is too limited in the context of causal sets and
the stages paradigm. Mathematicians generally prefer to work with vector
spaces whereas physicists are more concerned with particular subsets of
vector spaces, so we must extend our classification of vectors to
distinguish separable and entangled vectors. We will explain our terminology
starting with the separable sets.

Henceforth, we will assume that each subregister cannot itself be split in
any way, i.e., is an elementary subregister, such as a qubit.

Take any two subregisters $\mathcal{H}_{i},\mathcal{H}_{j}$, in $\left( \ref
{111}\right) $ such that $1\leqslant i,j\leqslant N$ and $i\neq j$. We
define the \emph{separation }$\mathcal{H}_{ij}$ of $\mathcal{H}_{[ij]}\equiv
\mathcal{H}_{i}\otimes \mathcal{H}_{j}$ to be the subset of $\mathcal{H}_{%
\left[ ij\right] }$ consisting of all separable elements, i.e.,
\begin{equation}
\mathcal{H}_{ij}\equiv \left\{ |\phi \rangle _{i}\otimes |\psi \rangle
_{j}:|\phi \rangle _{i}\in \mathcal{H}_{i},\;|\psi \rangle _{j}\in \mathcal{H%
}_{j}\right\} .
\end{equation}
By definition, we include the zero vector $0_{ij}$ of $\mathcal{H}_{\left[ ij%
\right] }$ in $\mathcal{H}_{ij}$ because this vector can always be written
as a trivially separable state, i.e.,
\begin{equation}
0_{ij}=0_{i}\otimes 0_{j}.
\end{equation}

The separation $\mathcal{H}_{ij}$ will be called a \emph{rank-2 separation}
and this generalizes to higher rank separations as follows. Pick an integer $%
k$ in the interval $\left[ 1,N\right] $ and then select $k$ different
elements $i_{1},i_{2},\ldots ,i_{k}$ of this interval.\ Then the \emph{%
rank-k separation }$\mathcal{H}_{i_{1}i_{2}\ldots i_{k}}$ is the subset of $%
\mathcal{H}_{\left[ i_{1}\ldots i_{k}\right] }\equiv \mathcal{H}%
_{i_{1}}\otimes \mathcal{H}_{i_{2}}\otimes \ldots \otimes \mathcal{H}%
_{i_{k}} $ given by
\begin{equation}
\mathcal{H}_{i_{1}i_{2}\ldots i_{k}}\equiv \left\{ |\psi _{1}\rangle
_{i_{1}}\otimes \ldots \otimes |\psi _{k}\rangle _{i_{k}}:|\psi _{a}\rangle
_{i_{a}}\in \mathcal{H}_{i_{a}},\;\;\;1\leqslant a\leqslant k\right\} .
\end{equation}
Every element of a rank-k separation has $k-$ factors. A rank-1 separation
of a subregister is by definition equal to that subregister and so we may
write
\begin{equation}
\mathcal{H}_{i}=\mathcal{H}_{[i]}.
\end{equation}

\

Now we are in a position to construct the \emph{entanglements}, which are
defined in terms of complements. Starting with the lowest rank possible, we
define the \emph{rank-2 entanglement }$\mathcal{H}^{ij}$ to be the
complement of $\mathcal{H}_{ij}$ in $\mathcal{H}_{\left[ ij\right] }$, i.e.,
\begin{equation}
\mathcal{H}^{ij}\equiv \mathcal{H}_{[ij]}\mathcal{-H}_{ij}=\left( \mathcal{H}%
_{[ij]}\mathcal{\cap H}_{ij}\right) ^{c}.
\end{equation}
Hence
\begin{equation}
\mathcal{H}_{[ij]}=\mathcal{H}_{ij}\cup \mathcal{H}^{ij}.
\end{equation}
Note that $\mathcal{H}_{ij}$ and $\mathcal{H}^{ij}$ are disjoint and $%
\mathcal{H}^{ij}$ does not contain the zero vector. An important aspect of
this decomposition is that neither $\mathcal{H}_{ij}$ nor $\mathcal{H}^{ij}$
is a vector space, as can be readily proved.

The generalization of the above to larger tensor product spaces is
straightforward but first it will be useful to extend our notation to
include the concept of \emph{separation product}. If $\mathcal{H}%
_{i}^{\prime }$ and $\mathcal{H}_{j}^{\prime }$ are arbitrary subsets of $%
\mathcal{H}_{i}$ and $\mathcal{H}_{j}$ respectively, where $i\neq j$, then
we define the separation product $\mathcal{H}_{i}^{\prime }\bullet \mathcal{H%
}_{j}^{\prime }$ to be the subset of $\mathcal{H}_{\left[ ij\right] }$ given
by
\begin{equation}
\mathcal{H}_{i}^{\prime }\bullet \mathcal{H}_{j}^{\prime }\equiv \left\{
|\psi \rangle \otimes |\phi \rangle :|\psi \rangle \in \mathcal{H}%
_{i}^{\prime },\;|\phi \rangle \in \mathcal{H}_{j}^{\prime }\right\} .
\end{equation}
This generalizes immediately to any sort of product. For example, we see
\begin{equation}
\mathcal{H}_{ij}=\mathcal{H}_{i}\bullet \mathcal{H}_{j}.
\end{equation}
The separation product is associative, commutative and cumulative, i.e.
\begin{align}
\left( \mathcal{H}_{i}\bullet \mathcal{H}_{j}\right) \bullet \mathcal{H}%
_{k}& =\mathcal{H}_{i}\bullet \left( \mathcal{H}_{j}\bullet \mathcal{H}%
_{k}\right) \equiv \mathcal{H}_{ijk}  \notag \\
\mathcal{H}_{ij}\bullet \mathcal{H}_{k}& =\mathcal{H}_{ijk},
\end{align}
and so on. The separation product can also be defined to include
entanglements. For example,
\begin{equation}
\mathcal{H}^{ij}\bullet \mathcal{H}_{k}=\left\{ |\phi \rangle _{ij}\otimes
|\psi \rangle _{k}:|\phi \rangle _{ij}\in \mathcal{H}^{ij},|\psi \rangle
_{k}\in \mathcal{H}_{k}\right\}
\end{equation}

A further notational simplification is to use a single $\mathcal{H}$ symbol,
using the vertical position of indices to indicate separations and
entanglements, and incorporating the separated product symbol $\bullet $
with indices directly. For example, the following are equivalent ways of
writing the same separation product of entanglements and separations:
\begin{equation}
\mathcal{H}_{23468}^{15\bullet 97}\equiv \mathcal{H}_{28\bullet 4\bullet
36}^{15\bullet 97}\equiv \mathcal{H}^{15}\bullet \mathcal{H}^{97}\bullet
\mathcal{H}_{28}\bullet \mathcal{H}_{4}\bullet \mathcal{H}_{36}.
\end{equation}
Associativity of the separation product applies to both separations and
entanglements. For example, we may write
\begin{equation}
\mathcal{H}_{ij}\bullet \mathcal{H}_{klm}\bullet \mathcal{H}^{rs}=\mathcal{H}%
_{ij\bullet klm}^{rs}=\mathcal{H}_{ijklm}^{rs},
\end{equation}
but note that whilst
\begin{equation}
\mathcal{H}_{ij}\bullet \mathcal{H}_{kl}=\mathcal{H}_{ij\bullet kl}=\mathcal{%
H}_{ijkl,}
\end{equation}
we have
\begin{equation}
\mathcal{H}^{ij}\bullet \mathcal{H}^{kl}\equiv \mathcal{H}^{ij\bullet
kl}\neq \mathcal{H}^{ijkl}.
\end{equation}
In practice, rank-3 and higher entanglements such as $\mathcal{H}^{ijkl}$
have to be defined in terms of complements, in the same way that $\mathcal{H}%
^{ij}$ is defined. For example, consider $\mathcal{H}_{[abc]}\equiv \mathcal{%
H}_{a}\otimes \mathcal{H}_{b}\otimes \mathcal{H}_{c}.$ We define
\begin{equation}
\mathcal{H}^{abc}\equiv \mathcal{H}_{[abc]}-\mathcal{H}_{abc}\cup \mathcal{H}%
_{a}^{bc}\cup \mathcal{H}_{b}^{ac}\cup \mathcal{H}_{c}^{ab}.
\end{equation}
Likewise, given $\mathcal{H}_{[abcd]}\equiv \mathcal{H}_{a}\otimes \mathcal{H%
}_{b}\otimes \mathcal{H}_{c}\otimes \mathcal{H}_{d}$ then
\begin{align}
\mathcal{H}^{abcd}& \equiv \mathcal{H}_{[abcd]}-\mathcal{H}_{abcd}\cup
\mathcal{H}_{ab}^{cd}\cup \mathcal{H}_{ac}^{bd}\cup \mathcal{H}%
_{ad}^{bc}\cup \mathcal{H}_{bc}^{ad}\cup \mathcal{H}_{bd}^{ac}\cup \mathcal{H%
}_{cd}^{ab}\cup  \notag \\
& \mathcal{H}_{a}^{bcd}\cup \mathcal{H}_{b}^{acd}\cup \mathcal{H}%
_{c}^{abd}\cup \mathcal{H}_{d}^{abc}\cup \mathcal{H}^{ab\bullet cd}\cup
\mathcal{H}^{ac\bullet bd}\cup \mathcal{H}^{ad\bullet bc}.
\end{align}
We will refer to sets such as $\mathcal{H}^{abc}$ as a rank-3 entanglement,
and so on. It is clear that in general, higher rank entanglements such as $%
\mathcal{H}^{abcd}$ in the above require a deal of filtering out of
separations from the original tensor product Hilbert space for their
definition to be possible, and this accounts partly for the fact that
entanglements are generally not as conceptually simple or intuitive as pure
separations. It is not surprising that classical mechanics is easier to
visualize than its quantum counterpart, because the former deals with
separations exclusively whilst the latter deals with separations and
entanglements.

We shall refer to the unique decomposition of a quantum register into the
union of disjoint separations and entanglements as the \emph{natural lattice}
$\frak{L}\left( \mathcal{H}\right) $ of $\mathcal{H}$, each element of which
will be referred to as a \emph{partition}. In general, partitions are
separation products of entanglements and separations of various ranks, such
that for each partition, the sum of the ranks of its factors equals the
number of quantum subregisters. The individual separations and entanglements
making up a partition will be called \emph{blocks}.

\

The relationships between separations and entanglements are subtle and will
be discussed in subsequent papers. The relationships between splits and
partitions are equally important. Although the number of partitions in the
natural lattice of a rank-$n$ quantum register is the same as the number of
splits, i.e., the $n^{th}$ Bell number, as can be readily proved, in
general, splits and partitions cannot coincide. Every factor in a split is a
vector space whereas no block in a partition is a vector space. Both splits
and partitions are essential ingredients in the construction of causal set
structure in our paradigm.

\

In the following, we shall use the above index notation to label the various
elements of entanglements and separations. For example, $\psi
_{123}^{456\;\bullet 78}$ is interpreted to be some element in $\mathcal{H}%
_{123}^{456\;\bullet 78h}$ and so on. With this notation we are entitled to
rewrite the vector $\psi _{123}^{456\bullet 78}$ in the factorized form
\begin{equation}
\psi _{123}^{456\;\bullet 78}=\psi _{1}\otimes \psi _{2}\otimes \psi
_{3}\otimes \psi ^{456}\otimes \psi ^{78},
\end{equation}
where $\psi _{1}\in \mathcal{H}_{1},\;\psi _{2}\in \mathcal{H}_{2}$, $\psi
_{3}\in \mathcal{H}_{3}$, $\psi ^{456}$ $\in \mathcal{H}^{456}$ and $\psi
^{78}\in \mathcal{H}^{78}$.

\

\begin{center}
{\large {\textbf{V. PROBABILITY AMPLITUDES, FAMILY STRUCTURE AND CAUSAL SETS}%
}}
\end{center}

We are now in a position to discuss causal sets proper. An important and
natural feature of the stages paradigm (see Appendix) is that the state of
the universe can change its factorizability as it jumps from $\Psi_{n}$ to $%
\Psi _{n+1}$ and this is the origin of family structure, as we shall
demonstrate.

To every stage $\Omega _{n}$ of the universe, we can assign a positive
integer $\mathcal{F}_{n}$, the \emph{current factorizability of the
universe. }This is just the number of factor states in $\Psi _{n}$ and we
may write
\begin{equation}
\Psi _{n}=\Psi _{n}^{1}\otimes \Psi _{n}^{2}\otimes \ldots \otimes \Psi
_{n}^{\mathcal{F}_{n}},
\end{equation}
where the subscript $n$ refers to exo-time (unphysical, i.e., unobservable,
external time) and the superscripts now label the various factors, rather
than denoting entanglements, there being $\mathcal{F}_{n}$ of these factors.
The various factor states \thinspace $\left\{ \Psi _{n}^{{}\alpha
}:1\leqslant {}\alpha \leqslant \mathcal{F}_{n}\right\} $ form a discrete
set referred to as the \emph{factor lattice }$\Lambda _{n}$.\ Although the
dimension of the total Hilbert space is fixed, the number of elements in the
factor lattice is time dependent and this forms the correct basis for a
discussion of quantum causal sets.

In the conventional causal set paradigm, the discussion is in terms of a
classical structure of sets related in various ways. We shall call such a
model a \emph{CCM }(classical causal model), whereas our paradigm will be
referred to as a \emph{QCM} (quantum causal model).

In a \emph{CCM}, there is a concept of \emph{internal temporality}, which
means that each element is born either to the future of, or is unrelated to,
all existing elements; that is, no element can arise to the past of an
existing element. In our $\emph{QCM,}$ every realized or potential stage is
the outcome of some actual or potential test, and cannot be regarded as in
the past of any realized or potential stage which is the origin of that
test. Therefore the stages paradigm also has built-in internal temporality.

In \emph{CCMs, }the \emph{irreflexive convention} states that an element of
a causal set does not precede itself. In our \emph{QCM}, this is another
expression of the consequences of the Kochen-Specker theorem \cite
{KOCHEN+SPECKER-67}, consistent with the notion that for quantum states,
classical values such as position and momentum do not exist prior to
measurement.

\

In causal set theory, a \emph{link} is an irreducible relation, i.e., one
not implied by other relations via transitivity$.$ Such a relation is also
referred to as a \emph{covering relation} in the mathematical literature. In
our \emph{QCM}, links are directly related to outcomes of tests and the
corresponding quantum amplitudes.

Consider the inner product between any two states in a quantum register.
Because of the natural lattice structure of any quantum register, each state
is in a specific partitio and, depending on the details of the partitions
concerned, the amplitude may or may not factorize. This is because factor
states can only take inner products in combinations which lie in the same
factor Hilbert space of some split of the total Hilbert space (quantum
``zipping"). The following example illustrates the point:

\begin{enumerate}
\item[\textbf{Example 1:}]  Consider the quantum register $\mathcal{H}_{%
\left[ 1\ldots 8\right]}.$ By inspection, the inner product of the states $%
\psi _{123}^{456\bullet 78}$ and $\phi _{145}^{23\bullet 678}$ takes the
factorized form
\begin{align}
\langle \phi _{145}^{23\bullet 678}|\psi _{123}^{456\bullet 78}\rangle &
=\langle \phi _{1}|\psi _{1}\rangle \langle \phi ^{23}|\psi _{23}\rangle
\langle \phi _{45}^{678}|\psi ^{456\bullet 78}\rangle  \notag \\
& =\langle \phi _{1}|\psi _{1}\rangle \langle \phi ^{23}|\left\{ |\psi
_{2}\rangle \otimes |\psi _{3}\rangle \right\} \\
& \times \left\{ \langle \phi ^{678}|\otimes \langle \phi _{4}|\otimes
\langle \phi _{5}|\right\} \left\{ |\psi ^{456}\rangle \otimes |\psi
^{78}\rangle \right\} ,  \notag
\end{align}
which cannot be simplified further.
\end{enumerate}

It is at this point that we find the natural and logical motivation for the
concept of \emph{family} in quantum causal set theory. The individual
elements of a family are none other than groups of entanglements and
separations associated with each separate factor in transition amplitudes.

Both splits and partitions are involved in such amplitudes, which can be
more readily appreciated when we represent them graphically. The graphical
notation we introduce here also turns out to be very useful in discussing
causal set dynamics. The convention is to represent each possible factor in
each state involved in an amplitude by a large circle, whilst common split
factors linking different but compatible blocks are represented by smaller
circles. Ket vectors represent earlier states and the diagram is drawn with
time running upwards. With this convention, the amplitude in Example 1 is
represented by Figure $1.$

\begin{center}
\begin{figure}[t]
\centerline{\epsfxsize=5.0in \epsfbox{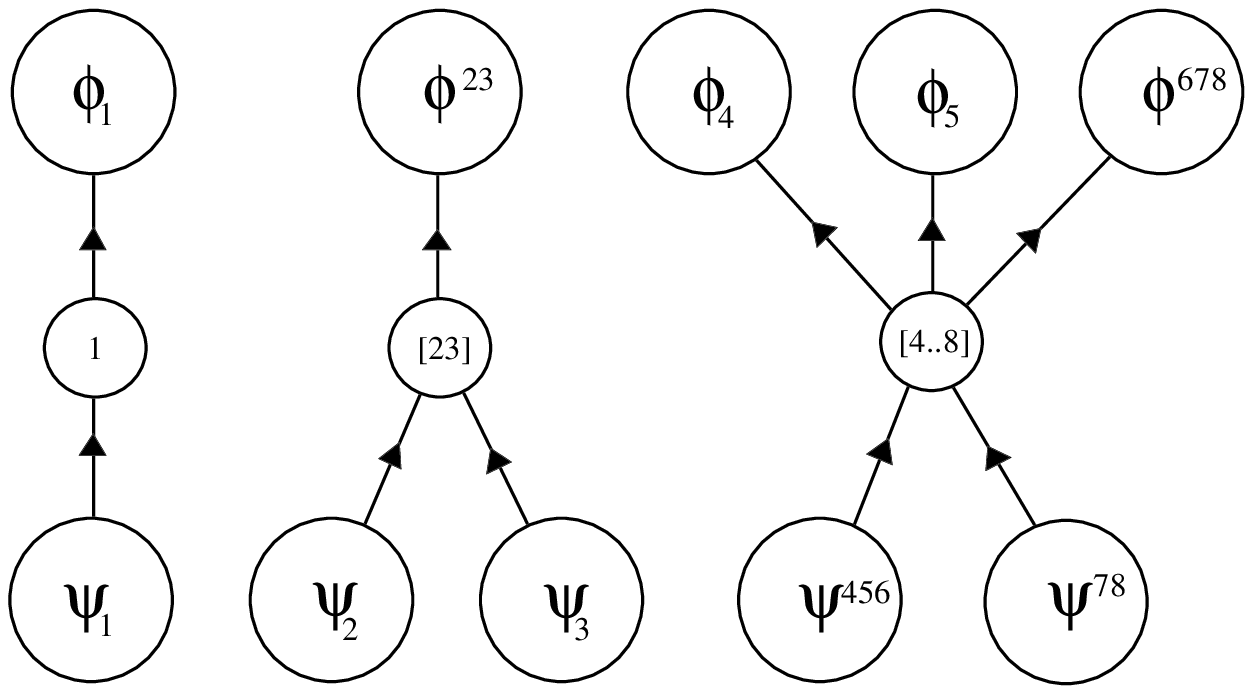}}
\caption{A typical quantum amplitude between states in a quantum register,
showing the appearance of family structure.}
\end{figure}
\end{center}

The next step is to incorporate the above to the stages paradigm. We shall
take it for granted that the total Hilbert space $\mathcal{H}_{\left[
1\ldots N\right] }$ is some vast tensor product of $N$ elementary
subregisters. Next, suppose that the current present state of the universe $%
\Psi _{n}$ has $k$ factors, i.e., $\Psi _{n}=\psi _{1}\otimes \psi
_{2}\otimes \ldots \otimes\psi_{k}$; each of which is a separation or
entanglement. Without loss of generality, we may take each of these factors
to have unit norm \ relative to the factor Hilbert space it lies in. Suppose
now that the next test of the universe $O_{n+1}$ is determined and that $%
\Theta $ is some normalized eigenstate of the corresponding observable $\hat{%
O}_{n+1}$. Then $\Theta $ is a possible candidate for $\Psi _{n+1}$, the
next state of the universe after $\Psi _{n}$. According to the principles
discussed in the Appendix, the probability of this outcome relative to $\Psi
_{n}$ is the conditional probability
\begin{equation}
P\left( \Psi _{n+1}=\Theta |\Psi _{n},\hat{O}_{n+1}\right) =|\langle \Theta
|\Psi _{n}\rangle |^{2}.
\end{equation}
Suppose further that $\Theta $ itself separates into $l$ factors, i.e. $%
\Theta =\theta _{1}\otimes \theta _{2}\otimes \ldots \otimes \theta _{l}$,
each of which is normalized to unity. If the pattern of subregisters
associated with the corresponding partitions in which $\Psi _{n}$ and $%
\Theta $ lie is such that $P\left( \Psi _{n+1}=\Theta |\Psi _{n},\hat{O}%
_{n+1}\right) $ itself factorizes into a number of factors, i.e.,
\begin{equation}
P\left( \Psi _{n+1}=\Theta |\Psi _{n},\hat{O}_{n+1}\right) =P_{1}P_{2}\ldots
P_{r},\;\;\;\;\;r\leqslant min(k,l),
\end{equation}
then each of these factors $P_{i}$ can be interpreted as a conditional
transition probability within a distinct family.

\

\begin{enumerate}
\item[\textbf{Example 2:}]  If in Example 1 we take $\Psi _{n}\equiv \psi
_{123}^{456\bullet 78}$ and $\Theta \equiv \phi _{145}^{23\bullet 678}$,
then $N=8,\;k=l=5$, $r=3$ and
\begin{equation}
P_{1}=|\langle \phi _{1}|\psi _{1}\rangle |^{2},\;\;\;P_{2}=|\langle \phi
^{23}|\psi _{23}\rangle |^{2},\;\;\;P_{3}=|\langle \phi _{45}^{678}|\psi
^{456\bullet 78}\rangle |^{2}.
\end{equation}
\end{enumerate}

From examples such as this one, we are led to give the following definition
of what we mean by family structure in the stages paradigm.

\begin{enumerate}
\item[\textbf{Definition 1:}]  Given a quantum register consisting of two or
more sub-registers, then in any quantum transition $|\Psi _{n}\rangle
\rightarrow |\Psi _{n+1}\rangle $, the number of families involved in that
transition is the number of factors in the transition amplitude $\langle
\Psi _{n+1}|\Psi _{n}\rangle $, as determined via the sub-register structure
of the register. Each of these factors identifies a unique family.

\item[\textbf{Comment 1:}]  Each factor in a transition amplitude involves a
distinct subset of the quantum subregisters making up the total Hilbert
space. Therefore, the factors collectively in such a transition amplitude
define a particular split of the total Hilbert space.
\end{enumerate}

Once a family has been identified, it is possible to define \emph{parents,
offspring} and \emph{siblings:}

\begin{enumerate}
\item[\textbf{Definition 2:}]  \ In a given family transition, all the
factors of the initial state of the family are \emph{parents}, whilst the
corresponding factors in the final state are \emph{offspring}, and are \emph{%
siblings} of each other if there are two or more such factors.
\end{enumerate}

In example 2, $\psi _{1}$ is the single parent of $\phi _{1}$, which has no
siblings. $\psi _{2}$ and $\psi _{3}$ \ are the parents of the entangled
state $\phi ^{23},$ and $\psi ^{456}$ and $\psi ^{78}$ are the parents of $%
\phi _{4},\phi _{5}$ and $\phi ^{678}$, which are siblings.

In general, extended sets of transitions, such as $\Psi _{n}\rightarrow \Psi
_{n+1}\rightarrow \Psi _{n+2}\rightarrow \ldots $, etc., will generate all
the properties of causal sets, such as grandparents, grandchildren, and
suchlike. Individual families may merge with other families or persist with
reasonably stable identities, depending on the specific transitions
involved. As we discuss in a later section, the details of the potential
transitions will be determined by the quantum tests involved.

Because we are dealing with quantum mechanics over a large rank quantum
register, there will be in general many possible outcomes of each test. Each
of these outcomes will then influence which future tests are chosen, so the
range of potential futures grows enormously the more jumps we consider. The
set of all alternative causal sets (potential futures) may be discussed in
terms of a grand causal set structure, analogous to the concept of poset
referred to in \cite{SORKIN-99}. It is here that the concept of entropy will
play a significant role. We will report on this elsewhere, noting that there
will in general be two contributions to the entropy related to a single
jump: the entropy associated with all the possible outcomes of any given
test, and the entropy associated with all the possible tests which could
occur with the given present.

Causal set structure emerges once we start to deal with more than two jumps.
To illustrate the sort of causal set structures which quantum registers can
generate, consider a Hilbert space $\mathcal{H}_{\left[ 123456\right] }$
over $6$ subregisters and the following sequence of normalized states:
\begin{equation}
\Psi ^{123456}\rightarrow \psi _{1}^{23\bullet 456}\rightarrow \theta
_{16}^{24\bullet 35}\rightarrow \eta _{4}^{12\bullet 356}\rightarrow \phi
^{12\bullet 34\bullet 56}\rightarrow \ldots
\end{equation}
From the factorization structure of the probabilities $P\left( \psi
_{1}^{23\bullet 456}|\Psi ^{123456}\right) \equiv |\langle \psi
_{1}^{23\bullet 456}|\Psi ^{123456}\rangle |^{2},$ etc., we can draw the
Hasse diagram structure shown in Figure 2.

\begin{center}
\begin{figure}[t]
\centerline{\epsfxsize=2.4in \epsfbox{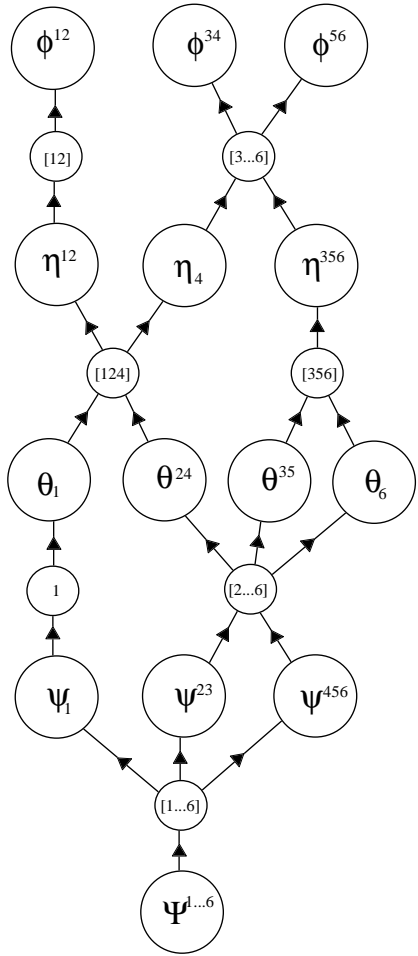}}
\caption{An example of causal set structure determined by successive states}
\end{figure}
\end{center}

In this diagram we see several key features worth commenting on.

\begin{enumerate}
\item  The initial state of the universe $\Psi ^{123456}$ is fully
entangled. Because in our paradigm separability is a marker of classicality,
the initial state has no classical attributes. It is not possible even to
think in terms of observers and systems at this stage;

\item  After the first jump, the universe has started to acquire
classicality, in the form of separability of its state. We refer to this as
the ``quantum Big Bang''. In our paradigm, the ``Big Crunch'' is synonymous
with a return to full entanglement;

\item  In the second jump, the factor $\psi _{1}$ changes to $\theta _{1}$,
with no connection with any other part of the universe. To all intents and
purposes, the universe appears to have split into two separate sub-universes
having nothing to do with each other. In a universe with very many
subregisters, it should be possible to use this sort of process to discuss
quantum black hole physics and the issue of information loss into the
interior of a black hole;

\item  If it happened to be the case that $\theta _{1}=\psi _{1}$, then to
all intents and purposes that part of the universe would appear to have its
local ``endo-time'' frozen whilst the rest of the universe evolved. This
constitutes what we call a local null test. Essentially, such a test leaves
the state that it is testing unchanged, i.e., the initial state happens to
be an eigenstate of the test. An example of this is the passage of an
electron prepared via the spin-up channel of one Stern-Gerlach device
through another identically orientated Stern-Gerlach device.

This mechanism has a number of important consequences. One of these is that
it permits a description of stage dynamics in terms of a local concept of
time, ``endo-time'', which unlike the external time used to label successive
stages of the universe can have physical significance, in that it relates
changes in factors of the state of the universe to each other. Endo-time is
non-integrable, unlike exo-time, because the number of physically
significant jumps any part of the universe undergoes depends on the details
of those jumps. In our paradigm, local null tests are expected to be the
origin of proper time in relativity (which is regarded here as an emergent
theory);

\item  Another consequence of null tests is that despite the absolute nature
of exo-time (which labels states of the universe), there is no reason to
suppose that endo-time is absolute. For instance, in Figure 2, if $\psi
_{1}=\theta _{1}$, we could regard $\psi _{1}$ as simultaneous with $\theta
^{24}$, $\theta ^{35}$ and $\theta _{6}$ without any contradictions arising.
Likewise, we could regard $\theta _{1} $ as simultaneous with $\psi ^{23}$
and $\psi^{456}$.

In the long run it should be possible to use local null tests and very large
rank quantum registers to explain the emergence of relativistic physics and
the equivalence of local inertial frames. This would require in addition a
careful account of the process physics of quantum measurement, none of which
is generally given in conventional theory. For instance, whilst it may be
meaningful to give an \emph{a posteriori} relativistically covariant
discussion of \emph{expectation values}, the same cannot be said of any
individual outcome of a single run of a quantum experiment, because a single
quantum outcome cannot be ``observed" by two different observers using
different equipment.

\item  A form of `lightcone'' structure can be seen to occur when we look at
the relationship between $\theta _{1}$ and $\eta ^{356}.$ These are
incomparable elements in the language of causal sets. The latter factor
looks as if it would be unchanged by any counterfactual change in the
former. However, we caution against taking this aspect of the Hasse diagram
too literally here. The diagram we are dealing with in Figure 2 involves
factors in amplitudes, which are known to have non-local correlations. What
is not shown in Figure 2 is another causal set, the one associated with the
tests which give the outcomes discussed here. If indeed we considered a
change in $\theta _{1}$, it is quite possible that subsequent tests did not
even have any factorizable outcomes, i.e. $\eta ^{356}$ need not exist as
part of a potential future of an altered $\theta _{1}.$

\item  Family structure per se is determined by the structure of successive
splits of the quantum register. The causal set structure related to this
aspect of the dynamics can be identified if the circles representing state
factors are replaced by lines, leaving the smaller circles representing
split factors. This shows more clearly how different families relate to each
other, rather than their individual family members. For example, Figure 2
gives the reduced diagram Figure 3.
\end{enumerate}

\begin{center}
\begin{figure}[t]
\centerline{\epsfxsize=2.4in \epsfbox{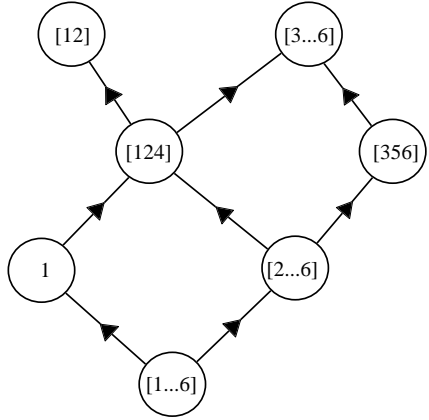}}
\caption{Reduced causal set structure involving successive splits in Figure
2.}
\end{figure}
\end{center}

\

\begin{center}
{\large {\textbf{VI. EINSTEIN LOCALITY AND CAUSAL SETS}} }
\end{center}

The appearance of family structures relating factors in successive states of
the universe, together with the very large number of factors observed in the
current epoch of the universe, opens the door to a discussion of metrical
and other classical concepts. Such a discussion is needed because emergent
space-time appears to have all the hallmarks of a classical
pseudo-Riemannian manifold, with an integral dimension, a Lorentzian
signature metric and curvature induced by the local presence of matter. A
particular feature of separability is distinguishability, that is, the
various factors of a separable state can be identified directly with the
corresponding quantum subregisters concerned. These registers are regarded
as distinct, and therefore, this may be regarded as the origin of
classicality (i.e., the ability to distinguish objects). The same cannot be
said of entangled states. In classical mechanics, for instance, it is well
known that states of systems cannot be entangled, because their properties
have to be well defined.

We envisage that many intricate patterns and hierarchies of patterns of
related factors could persist in approximate detail as the universe jumps
from stage to stage, thereby creating the appearance of a universe with
semi-classical structures, analogous to various patterns seen in Conway's
``Game of Life", for example. These structures could have approximate
relationships describable in terms of space, distance and other classical
constructs by endo-physical observers, themselves described by such patterns
of factors. Underlying this description would be the counting procedures
used by such endo-physical observers to register approximate estimates of
jumps (giving rise to measurements of emergent time) and approximate
estimates of family relationships (giving rise to emergent concepts of
space). Exactly how emergent spacetime and matter could arise from causal
set familial relationships and persistence is an enormous problem reserved
for the future.

The observation of quantum correlations creates a problem for such an
emergent picture, however. Although much of physics appears to respect
Einstein locality \cite{PERES:93}\ as far as causality is concerned, the
non-local superluminal quantum correlations observed in \emph{EPR} type
experiments appear incompatible with the principles of relativity. This is
consistent with the notion that there are really two different causal
mechanisms involved, each with its own variety of ``distance''
relationships. One mechanism is involved with Einstein locality and all that
it implies, such as a maximum speed (the speed of light) for the propagation
of physical information such as energy and momentum, whereas the other
mechanism is responsible for the transmission of quantum correlations, which
appear to occur with no limitation of speed in any frame \cite{SCARANI-00}.
The problem for conventional physics based on Lorentzian manifolds is that
it cannot ``explain'' the latter mechanism. If Einstein locality is
synonymous with classical Lorentzian manifold structure, and if this
structure is emergent, then it seems reasonable to interpret quantum
correlations as a signal that there is a pre-geometric (or pre-emergent)
structure underlying the conventional spacetime paradigm.

In the stages paradigm, there naturally occurs two different components
involved in the dynamical evolution. These are the tests and the outcomes of
those tests respectively and the properties and characteristics of each of
these components differ. Our hypothesis is that the tests are responsible
for the appearance of Einstein locality, whereas it is the outcomes which
display non-local effects such as quantum correlations.

Conventional quantum field theory is consistent with this point of view.
There too there are two aspects to the dynamics, i.e., the quantum field
operators out of which the dynamical observables are constructed and the
quantum states. For various technical reasons, the Heisenberg picture is
generally the one employed in quantum field theory. In this picture, states
are frozen in time between state preparation and measurement, whilst
dynamical evolution is locked into the field operators. In this picture,
which happens also to be the best picture to discuss the stages paradigm,
quantum field operators satisfy classical equations of motion for their
evolution over a classical spacetime, i.e., they obey operator Heisenberg
equations of motion. This does not imply that super-luminal quantum
correlations cannot take place, because such correlations relate to the
properties of quantum states and not to the observables (the operators of
physical interest) of the theory. Local observable densities such as energy
and momentum density operators satisfy microscopic causality \cite
{BJORKEN+DRELL:65B}, \ which means that they have commutators which vanish
at relative spacelike intervals, even if the local fields out of which they
are constructed do not. \ This guarantees that Einstein locality holds as
far as the corresponding classical variables are concerned.

The relationship between quantum field theory and the stages paradigm is
even stronger. In scattering quantum field theory for example, there is the
same structure of state preparation, test and outcome as in the stages
paradigm. First an ``in'' state $|\Psi \rangle _{in}$ is prepared at what
amounts to the remote past. In that regime, the ``in'' state is taken to be
a many-particle eigenstate of some beam preparation apparatus. This
apparatus will be associated with some preferred basis set $\mathcal{B}_{in}$%
, one element of which is selected to be $|\Psi \rangle _{in}.$ The system
is then left alone until at what amounts to infinite future time, it is
tested against what is equivalent to a preferred basis $\mathcal{B}_{out}$
of free particle states, the ``out'' states. Because an understanding of the
Hilbert space of possible states in fully interacting quantum field theory
is virtually non-existent, the Heisenberg picture is invariably the one used
in scattering theory. The traditional vehicle for calculation is the
S-matrix formalism \cite{BJORKEN+DRELL:65B,GASIOROWICZ:68}, which gives the
transition probabilities generated by some unitary transformation of the
final state basis $\mathcal{B}_{out}$ relative to the initial (preparation)
state basis $\mathcal{B}_{in}.$ In the Heisenberg picture, it is not the
case that the state being tested changes unitarily in time. Rather, time is
something associated with the tests constructed by the observer.

\

\begin{center}
{\large {\textbf{VII. SKELETON SETS}} }
\end{center}

In the stages paradigm, the tests involved in jumps are not arbitrary but
are represented by specific operators determined by as yet unknown dynamical
principles. In particular, these operators are assumed to be Hermitian,
because the principles of standard $QM$ are based on such operators. In
general, if we are dealing with a finite dimensional Hilbert space of
dimension (say) $n$, then we can find \ $n^{2}$ independent operators \cite
{PERES:93} out of which we can build all possible Hermitian operators. This
leads us to a discussion of skeleton sets of operators.

To explain the notion of a skeleton set of operators, consider the simple
example of a qubit register, i.e., a two-dimensional Hilbert space $\mathcal{%
H}_{A}$, where the subscript $A$ labels the qubit, in anticipation of a
generalization to a many-qubit register. Given a preferred basis $\mathcal{B}%
_{A}\equiv \left\{ |1\rangle _{A},|2\rangle _{A}\right\} $ for $\mathcal{H}%
_{A}$ there are four Hermitian operators $\hat{\sigma}_{A}^{\mu }:\mu
=0,1,2,3$ which may be used to construct any Hermitian operator on $\mathcal{%
H}_{A}.$ Here $\hat{\sigma}_{A}^{0}$ is the identity operator on $\mathcal{H}%
_{A}$ and the $\hat{\sigma}_{A}^{i}:i=1,2,3$ are three operators analogous
to the Pauli matrices, satisfying the product rules
\begin{align}
\hat{\sigma}_{A}^{i}\hat{\sigma}_{A}^{j}& =\delta _{ij}\hat{\sigma}%
_{A}^{0}+i\varepsilon _{ijk}\hat{\sigma}_{A}^{k},  \notag \\
\hat{\sigma}_{A}^{i}\hat{\sigma}_{A}^{0}& =\hat{\sigma}_{A}^{0}\hat{\sigma}%
_{A}^{i}=\hat{\sigma}_{A}^{i},
\end{align}
where we use the summation convention with small Greek and small Latin
symbols, but not on the large Latin indices which label qubits. These rules
may be written in the more compact form
\begin{equation}
\hat{\sigma}_{A}^{\mu }\hat{\sigma}_{A}^{\nu }=c_{\;\;\;\alpha }^{\mu \nu }%
\hat{\sigma}_{A}^{\alpha },
\end{equation}
where the $c_{\;\;\;\alpha }^{\mu \nu }$ are given by
\begin{align}
c_{\;\;\mu }^{0\nu }& =c_{\;\;\;\mu }^{\nu 0}\equiv \delta _{\nu \mu },
\notag \\
c_{\;\;0}^{ij}& \equiv \delta _{ij} \\
c_{\;\;k}^{ij}& \equiv i\varepsilon _{ijk}.  \notag
\end{align}
In the standard basis $\frak{B}_{A}$ we may represent these operators by
\begin{equation}
\hat{\sigma}_{A}^{\mu }\equiv \sum_{i,j=1}^{2}|i\rangle _{A}[\sigma
_{A}^{\mu }]_{ij\;A}\langle j|,  \label{398}
\end{equation}
where the matrices $[\sigma _{A}^{\mu }]$ are defined by
\begin{align}
\lbrack \sigma _{A}^{0}]& \equiv \left[
\begin{array}{cc}
1 & 0 \\
0 & 1
\end{array}
\right] ,\;\;\;\;[\sigma _{A}^{1}]\equiv \left[
\begin{array}{cc}
0 & 1 \\
1 & 0
\end{array}
\right]  \notag \\
\lbrack \sigma _{A}^{2}]& \equiv \left[
\begin{array}{cc}
0 & -i \\
i & 0
\end{array}
\right] ,\;\;\;[\sigma _{A}^{3}]\equiv \left[
\begin{array}{cc}
1 & 0 \\
0 & -1
\end{array}
\right] .\;\;
\end{align}

The four operators $\hat{\sigma}_{A}^{\mu }$ will be called a \emph{skeleton
set} for $\mathcal{H}_{A}$ and denoted by $\mathcal{S}_{A}.$ The
significance of this set arises from the fact that an arbitrary linear
combination of elements of $\mathcal{S}_{A}$ with real coefficients is a
Hermitian operator. Moreover, any Hermitian operator on $\mathcal{H}_{A}$
can be uniquely represented as a real linear combination of elements of the
skeleton set. The skeleton set $\mathcal{S}_{A}$ therefore forms a basis for
the real vector space $\mathbb{H(}\mathcal{H}_{A})$ of (linear) Hermitian
operators on $\mathcal{H}_{A}.$ In addition, this vector space is closed
under multiplication based on the rule (\ref{398}) and therefore forms an
(linear) algebra.

Going further, consider a quantum register $\mathcal{H}$ consisting of $N$
qubits, given by the tensor product
\begin{equation}
\mathcal{H}\equiv \mathcal{H}_{1}\otimes \mathcal{H}_{2}\otimes \ldots
\otimes \mathcal{H}_{N}.
\end{equation}
We define a skeleton set $\mathcal{S}$ for $\mathcal{H}$ by the direct
product
\begin{equation}
\mathcal{S}\equiv \{\hat{\sigma}_{1}^{\mu _{1}}\otimes \hat{\sigma}_{2}^{\mu
_{2}}\otimes \ldots \otimes \hat{\sigma}_{N}^{\mu _{N}}:\;\;\;0\leqslant \mu
_{1},\mu _{2},\ldots ,\mu _{N}\leqslant 3\}.  \label{112}
\end{equation}
These are all Hermitian operators on $\mathcal{H}$ and form a basis for the
real vector space $\mathbb{H(\mathcal{H})}$ of Hermitian operators on $%
\mathcal{H}$. An arbitrary element $\hat{O}$ of $\mathbb{H(\mathcal{H})}$ is
of the form
\begin{equation}
\hat{O}\equiv O_{\mu _{1}\mu _{2}\ldots \mu _{N}}\hat{\sigma}_{1}^{\mu
_{1}}\otimes \hat{\sigma}_{2}^{\mu _{2}}\otimes \ldots \otimes \hat{\sigma}%
_{N}^{\mu _{N}},
\end{equation}
where we sum over the small Greek indices and the coefficients $O_{\mu
_{1}\mu _{2}\ldots \mu _{N}}$ are all real. Moreover, multiplication of
elements of $\mathbb{H(\mathcal{H})}$ is defined in a straightforward way.
We have
\begin{align}
\hat{A}\hat{B}& \equiv \left\{ A_{\mu _{1}\mu _{2}\ldots \mu _{N}}\hat{\sigma%
}_{1}^{\mu _{1}}\otimes \hat{\sigma}_{2}^{\mu _{2}}\otimes \ldots \otimes
\hat{\sigma}_{N}^{\mu _{N}}\right\} \left\{ B_{\nu _{1}\nu _{2}\ldots \nu
_{N}}\hat{\sigma}_{1}^{\nu _{1}}\otimes \hat{\sigma}_{2}^{\nu _{2}}\otimes
\ldots \otimes \hat{\sigma}_{N}^{\nu _{N}}\right\}  \notag \\
& =A_{\mu _{1}\mu _{2}\ldots \mu _{N}}B_{\nu _{1}\nu _{2}\ldots \nu
_{N}}c_{\;\;\;\;\;\alpha _{1}}^{\mu _{1}\nu _{1}}c_{\;\;\;\;\;\alpha
_{2}}^{\mu _{2}\nu _{2}}\ldots c_{\;\;\;\;\;\alpha _{N}}^{\mu _{N}\nu _{N}}%
\hat{\sigma}_{1}^{\alpha _{1}}\otimes \hat{\sigma}_{2}^{\alpha _{2}}\otimes
\ldots \otimes \hat{\sigma}_{N}^{\alpha _{N}},
\end{align}
which is some element of $\mathbb{H(\mathcal{H})}$, which means that $%
\mathbb{H(\mathcal{H})}$ is an algebra \cite{HOWSON:72}\ over the real
number field.

Given a large $N$ qubit register $\mathcal{H}$, then in the stages paradigm,
dynamical evolution on the pre-geometric level involves a succession of
tests $O_{n}$ and associated outcomes $\Psi _{n}$. In this paradigm, the
total Hilbert space $\mathcal{H}$ is fixed, in contrast to some other models
\cite{MARKOPOULOU-00,ZIZZI-01}. Assuming $N$ is extremely large and finite,
then there will be a natural skeleton set $\mathcal{S}$ given in (\ref{112})
which permits a decomposition of each test, i.e., we may write
\begin{equation}
\hat{O}_{n}=C_{\mu _{1}\mu _{2}\ldots \mu _{N}}^{n}\hat{\sigma}_{1}^{\mu
_{1}}\otimes \hat{\sigma}_{2}^{\mu _{2}}\otimes \ldots \otimes \hat{\sigma}%
_{N}^{\mu _{N}},  \label{113}
\end{equation}
where the coefficients $C_{\mu _{1}\mu _{2}\ldots \mu _{N}}^{n}$ are all
real. Here and elsewhere we shall use the summation convention.

We have already discussed the analogy between the behaviour of tests in the
stages paradigm and operators in quantum field theory. It may be the case
that for a given $n$, the set of coefficients $\left\{ C_{\mu _{1}\mu
_{2}\ldots \mu _{N}}^{n}\right\} $ is such that the right hand side in (\ref
{113}) factorizes, i.e., we may write
\begin{equation}
\hat{O}_{n}=\hat{A}\otimes \hat{B},
\end{equation}
where $\hat{A}$ is an operator acting on one factor subspace $\mathcal{H}%
_{A} $ of \ $\mathcal{H}$ and $\hat{B}$ acts on the other factor subspace $%
\mathcal{H}_{B}$. Together, $\mathcal{H}_{A}$ and $\mathcal{H}_{B}$ give a
bi-partite factorization or split of $\mathcal{H}$ \cite{JAROSZKIEWICZ-03A},
i.e.
\begin{equation}
\mathcal{H}=\mathcal{H}_{A}\otimes \mathcal{H}_{B}.
\end{equation}

We now discuss the necessary and sufficient conditions for a Hermitian
operator on $\mathcal{H}$ to factorize with respect to the basic skeleton
set.

\begin{enumerate}
\item[\textbf{Theorem 1:}]  Let $\mathcal{H}\equiv \mathcal{H}_{1}\otimes
\mathcal{H}_{2}$ be an $2$-qubit quantum register with standard basis
\begin{equation}
\mathcal{B}\equiv \left\{ |i_{1}\rangle _{1}\otimes |i_{2}\rangle
_{2}:1\leqslant i_{1},i_{2}\leqslant 2\right\}
\end{equation}
and standard skeleton set $\mathcal{S}\equiv \left\{ \hat{\sigma}_{1}^{\mu
_{1}}\otimes \hat{\sigma}_{2}^{\mu _{2}}:0\leqslant \mu _{1},\mu
_{2}\leqslant 3\right\} $ and let $\mathbb{H(\mathcal{H})}$ be the set of
Hermitian operators on $\mathcal{H}$. Then an arbitrary element $\hat{O}\in
\mathbb{H(\mathcal{H})}$, given by an expansion of the form
\begin{equation}
\hat{O}=c_{\mu \nu }\hat{\sigma}_{1}^{\mu }\otimes \hat{\sigma}_{2}^{\nu
},\;\;\;c_{\mu \nu }\in \mathbb{R},  \label{exp}
\end{equation}
factorizes with respect to $\mathcal{S}$ if and only if the coefficients $%
c_{\mu \nu }$ satisfy the \emph{micro-singularity}\textbf{\ }condition
\begin{equation}
c_{\mu \nu }c_{\alpha \beta }=c_{\mu \beta }c_{\alpha \nu }  \label{micro}
\end{equation}
for all values of the indices.

\item[\textbf{Proof:}]  $\Rightarrow :$ If an operator $\hat{O}\in \mathbb{H(%
\mathcal{H})}$ factorizes relative to $\mathcal{S}$ then we may write
\begin{align}
\hat{O}& =\left( a_{\mu }\hat{\sigma}_{1}^{\mu }\right) \otimes \left(
b_{\nu }\hat{\sigma}_{2}^{\nu }\right)  \notag \\
& =\left( a_{\mu }b_{\nu }\right) \hat{\sigma}_{1}^{\mu }\otimes \hat{\sigma}%
_{2}^{\nu },
\end{align}
which means we take $c_{\mu \nu }=a_{\mu }b_{\nu }$ \ for all values of the
indices, and these clearly satisfy the micro-singularity condition (\ref
{micro}).

$\Leftarrow :$ Suppose $\hat{O}\in \mathbb{H(\mathcal{H})}$ such that the
coefficients of its expansion (\ref{exp}) satisfy the micro-singularity
condition (\ref{micro}). Without loss of generality we may assume $\hat{O}$
is not the zero operator. Therefore, there is at least one coefficient $%
c_{\alpha \beta }$ in the expansion (\ref{exp}) which is non-zero. Hence we
may write
\begin{align}
c_{\alpha \beta }\hat{O}& =c_{\alpha \beta }c_{\mu \nu }\hat{\sigma}%
_{1}^{\mu }\otimes \hat{\sigma}_{2}^{\nu }  \notag \\
& =c_{\alpha \nu }c_{\mu \beta }\hat{\sigma}_{1}^{\mu }\otimes \hat{\sigma}%
_{2}^{\nu },\;\;\;using\;\;(\ref{micro}) \\
& =(c_{\mu \beta }\hat{\sigma}_{1}^{\mu })\otimes \left( c_{\alpha \nu }\hat{%
\sigma}_{2}^{\nu }\right) ,  \notag
\end{align}
which proves $\hat{O}$ is separable with respect to $\mathcal{S}$. This
result generalizes to more general sub-registers than just qubits.
\end{enumerate}

There are fundamental differences between the concept of entanglement of
states (for which a similar micro-singularity theorem holds \cite
{JAROSZKIEWICZ-03A}) and the ``entanglement'' of Hermitian operators. First,
the former involves the complex numbers whereas the latter involves the
reals. Quantum probabilities are extracted by taking the square moduli of
inner products of states, a process which generates the phenomenon of
quantum interference, an inherently quantum effect. No such phenomenon
arises with Hermitian operators, where at best only products of operators
are ever constructed (the operators form a ring). In general, there seems to
be no physically motivated concept of inner product on the space $\mathbb{H(%
\mathcal{H})}$ of Hermitian operators over a Hilbert space, and no
corresponding probability interpretation, although the stages paradigm
emphasizes the possibility that perhaps there should be such a thing.

The difference between states and observables should manifest itself in the
sort of causal sets they are associated with. In quantum mechanics
generally, states satisfy the principle of superposition, which leads to
non-local and non-classical consequences at odds with classical relativity,
whereas observables tend to have classical analogues which satisfy Einstein
locality and obey classical equations of motion consistent with relativity.
We recall that canonical quantization is the standard procedure of replacing
classical variables with their quantum operator counterparts. In standard
quantum mechanics, those operators usually satisfy the same causal relations
as their classical analogues. For example, those operators identified as
observables should commute at spacelike separations.

In the stages paradigm, we do not yet know the rules the universe uses for
choosing tests. However, because those tests should correspond to standard
observables in the appropriate emergent limits, one way of ensuring this is
if the causal set structure associated with successive tests follows
patterns analogous to classical cellular automata. A particular feature of
automata based on nearest neighbour interactions is the appearance of zones
of causal influence looking very much like lightcones in relativity and
fully consistent with Einstein causality.

In support of this line of argument, we refer to Theorem 4, proved in the
next section, which says that separability of a test necessarily implies
separability of outcome. Because separability is a necessary attribute of
classical space, Theorem 4 implies that separability of the observables
begins to drive classicality in the states, as discussed in Example 3.

Another difference between tests and states is that in general, for a fixed
dimension $N$ of the quantum register $\mathcal{H}$, the dimensionality of
the vector space $\mathbb{H(\mathcal{H})}$ of Hermitian operators is $N^{2}$%
, which means that the operators have a richer structure in terms of their
separability and entanglement than does the corresponding set of states.

Despite their expected differences, however, the relationship between the
separability and entanglement properties of states and of operators is a
deep and important one which is discussed next.

\

\begin{center}
{\large {\textbf{VIII. EIGENVALUES}} }
\end{center}

The question of eigenvalues of operators is a fundamental one, because the
spectrum of a Hermitian operator is directly associated with physical
information. This is also related to the concepts of separability and
entanglement of operators and to the important issue of \emph{preferred bases%
}.

In the following we shall assume all Hilbert spaces are finite dimensional
and make references to the following terms:

A \emph{degenerate} operator is a Hermitian operator with at least two
linearly independent eigenstates with identical eigenvalues.

A \emph{weak} operator is a Hermitian operator which is either degenerate or
at least one of its eigenvalues is zero.

A \emph{strong} operator is a Hermitian operator which is not weak; that is,
none of its eigenvalues are zero and all its eigenvalues are distinct.

\begin{enumerate}
\item[\textbf{Comment 2:}]  Degeneracy of eigenvalues represents a certain
loss of information, in that different eigenstates with the same eigenvalue
cannot be physically separated by the apparatus concerned. In this sense,
eigenvalues are not important in absolute terms \emph{per se}. What is
important is the knowledge that one eigenstate is distinguishable from
another, and that is why physicists generally try to construct tests
represented by non-degenerate operators.

\item[\textbf{Theorem 2:}]  All the normalized eigenstates of a strong
operator $\hat{O}$ form a unique, orthonormal basis set known as the \emph{%
preferred basis} relative to $\hat{O}$, denoted by $\frak{B}(\hat{O})$. The
number of these eigenstates equals the dimension of the Hilbert space on
which $\hat{O}$ acts.

\item[\textbf{Proof:}]  This is an extension of standard theory, the only
difference being that we deal with strong operators rather than
non-degenerate operators.

\item[\textbf{Comment 3:}]  Projection operators are not strong.

\item[\textbf{Comment 4:}]  Strong operators are important in the context of
tensor product registers because weak operators are in themselves
insufficient to determine a preferred basis, except in the particular case
of a single qubit system, which is of no interest here.
\end{enumerate}

Suppose now we have a tensor product Hilbert space $\mathcal{H}_{[12]}\equiv
\mathcal{H}_{1}\otimes \mathcal{H}_{2}$ and suppose $\hat{O}_{1}$ $\in
\mathbb{H}\left( \mathcal{H}_{1}\right) $ and $\hat{O}_{2}$ $\in \mathbb{H}%
\left( \mathcal{H}_{2}\right) $. Then the tensor product operator $\hat{O}%
_{12}\equiv \hat{O}_{1}\otimes \hat{O}_{2}$ is a separable element of $%
\mathbb{H}\left( \mathcal{H}_{\left[ 12\right] }\right) $. We observe the
following:

\begin{enumerate}
\item[i)]  If either $\hat{O}_{1}$ or $\hat{O}_{2}$ is weak (as far as their
eigenvalues with respect to their respective Hilbert spaces are concerned),
then $\hat{O}$ is necessarily degenerate.

\item[ii)]  If $\hat{O}_{1}$ and $\hat{O}_{2}$ are both strong, then $\hat{O}%
_{12}\equiv \hat{O}_{1}\otimes \hat{O}_{2}$ could be either strong or weak.

\item[\textbf{Comment 5:}]  Every element of the skeleton set $\left( \ref
{112}\right) $\ associated with an $n-$qubit register is weak for $n>1.$

\item[\textbf{Definition 3:}]  Let $\mathcal{S}\equiv \left\{ a,b,\ldots
,z\right\} $ be a finite set of real numbers. Then $\mathcal{S}$ is \emph{%
strong }if the elements are distinct and none is zero.

\item[\textbf{Definition 4:}]  The \emph{pair-wise product }$\mathcal{ST}$ \
of two finite real sets $\mathcal{S}\equiv \left\{ a,b,\ldots ,z\right\} ,%
\mathcal{T}\equiv \left\{ A,B,\ldots ,Z\right\} $ is the set of all products
of the elements of $\mathcal{S}$ with the elements of $\mathcal{T}$, i.e.,
\begin{equation}
\mathcal{ST}\equiv \left\{ aA,aB,\ldots ,zZ\right\} .
\end{equation}

\item[\textbf{Theorem 3:}]  A direct product $\hat{O}_{12}\equiv \hat{O}%
_{1}\otimes \hat{O}_{2}$ of two strong operators is strong if and only if
the pair-wise product of their corresponding spectra is strong.

Conversely, if the tensor product of two Hermitian operators is strong, then
each of these operators must also be strong.

\item[\textbf{Proof:}]  This follows from the fact that an operator is
strong if and only if its spectrum is strong, and from the observation that
the spectrum of a tensor product operator is the pair-wise product of their
respective spectra.
\end{enumerate}

This leads to the following theorem which has important implications for the
physics of entanglement:

\begin{enumerate}
\item[\textbf{Theorem 4:}]  (The fundamental theorem) All the eigenstates of
a separable strong operator operators are separable. Conversely, entangled
states can be produced only by entangled operators.

\item[\textbf{Proof:}]  Let $\mathcal{H}_{1}$ and $\mathcal{H}_{2}$ be two
Hilbert spaces of dimension $d_{1},d_{2}$ respectively and let $\hat{O}%
_{1}\in \mathbb{H}\left( \mathcal{H}_{1}\right) $ and $\hat{O}_{2}\in
\mathbb{H}\left( \mathcal{H}_{2}\right) .$ Then $\hat{O}_{12}\equiv \hat{O}%
_{1}\otimes \hat{O}_{2}\in \mathbb{H}\left( \mathcal{H}_{\left[ 12\right]
}\right) $ is a separable operator. Given that $\hat{O}$ is strong, then the
following must be true:

$i)$ By Theorem $2,$ $\hat{O}_{12}$ has precisely $d_{1}d_{2}$ distinct
eigenstates;

$ii)$ By Theorem $3,$ $\hat{O}_{1}$ and $\hat{O}_{2}$ are each necessarily
strong.

From $ii)$, we can find a unique orthonormal basis $\frak{B}_{1}\equiv
\left\{ u_{1},\ldots ,u_{d_{1}}\right\} $ for $\mathcal{H}_{1}$ consisting
of the distinct eigenstates of $\hat{O}_{1}.$ Likewise, we can find a unique
orthonormal basis $\frak{B}_{2}\equiv \left\{ v_{1},\ldots
,v_{d_{2}}\right\} $ for $\mathcal{H}_{2}$ consisting of the distinct
eigenstates of $\hat{O}_{2}.$ Now consider the product states $u_{i}\otimes
v_{j},$ $1\leqslant i\leqslant d_{1},1\leqslant j\leqslant d_{2}$. There are
exactly $d_{1}d_{2}$ such products, they are all separable, and each of
these is an eigenstate of $\hat{O}_{12}$, as can be readily proved. But
according to $i)$, $\hat{O}_{12}$ has only $d_{1}d_{2}$ distinct
eigenstates. Therefore, all the eigenstates of $\hat{O}_{12}$ are separable.

\item[\textbf{Comment 6:}]  This result immediately generalizes to tensor
products of $n\geqslant 2$ operators. It leads to the fundamental conclusion
that entangled states can be the outcomes of entangled operators only, a
fact which has significant implications in physics. Physics laboratories
which are not in any way correlated cannot create states entangled relative
to those laboratories.
\end{enumerate}

We may extend our graphical notation to include observables. Each completely
entangled observable will be denoted by a square. A separable operator with $%
k$ factors is represented by $k$ squares aligned left to right. Lines with
arrow going upwards and into such squares represent incoming prepared factor
states whilst outgoing lines running upwards represent outcomes.

With this convention, then Theorem 4 states that diagrams such as Figure
4(a) are forbidden whereas processes represented by Figure 4(b) are
permitted.

\begin{center}
\begin{figure}[t]
\centerline{\epsfxsize=3.0in \epsfbox{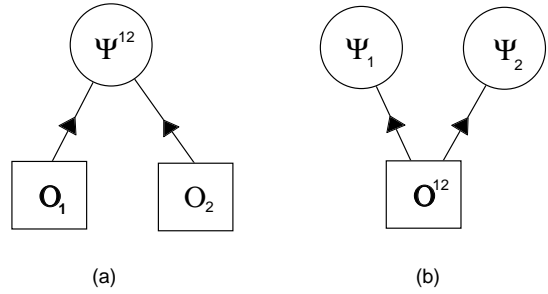}}
\caption{By Theorem 4, diagrams such as (a) are forbidden whereas those such
as (b) are not.}
\end{figure}
\end{center}

\newpage

\begin{center}
\textbf{Time reversal}
\end{center}

The stages paradigm has a built in irreversibility of time, because the
process of state preparation, test and outcome cannot in general be reversed
without altering the test. Suppose $\Psi$ is a prepared state, being an
outcome of test $\hat{A}$. Now consider a subsequent test $\hat{B}$ of $\Psi
$. If $\Theta $ is an eigenstate of $\hat{B}$ then the conditional
probability $P\left( \Theta |\Psi ,\hat{B}\right) $ of this outcome is given
by the standard Born rule
\begin{equation}
P\left( \Theta |\Psi ,\hat{B}\right) =|\langle \Theta |\Psi \rangle |^{2}.
\end{equation}
The right hand side is invariant to the interchange of $\Psi $ and $\Theta $%
, but the left hand side is not invariant to this change in general, because
$\Psi $ need not be an eigenstate of $\hat{B}$.

If we represent this sequence of events diagramatically, as in Figure 5a, it
is clear that merely interchanging the direction of the arrows need not be
physically meaningful. Instead, the sequence of tests has to be reversed as
well, as in Figure 5b, and then this gives the correct time-reversal
equality
\begin{equation}
P\left( \Psi |\Theta ,\hat{A}\right) =P\left( \Theta |\Psi ,\hat{B}\right) ,
\end{equation}
provided test $\hat{A}$ is a legitimate test of state $\Theta $, which
depends on the dynamics.

\begin{center}
\begin{figure}[t]
\centerline{\epsfxsize=2.0in \epsfbox{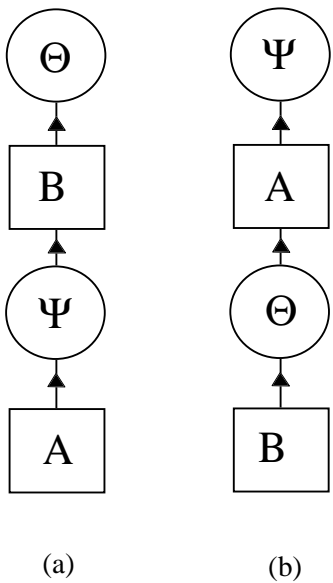}}
\caption{(b) is the time reversal of (a)}
\end{figure}
\end{center}

\newpage

\begin{center}
{\large {\textbf{IX. \ PHYSICAL EXAMPLES}}}
\end{center}

An important fact that has physical consequences is that entangled operators
can have entangled eigenstates \textbf{and} separable eigenstates, but it is
not true the other way around, according to Theorem $4.$ Factorizable strong
operators cannot have entangled eigenstates. Moreover, it is possible for an
entangled operator to have only separable eigenstates.

Given that there exists a structure of separations and entanglements for
observables (elements of $\mathbb{H}\left( \mathcal{H}\right) )$ relative to
skeleton sets of operators, then it is meaningful to talk about separable
and entangled observables \emph{per se. }Physically these concepts make
sense. It is possible to construct an experiment which prepares a
factorizable state, with each factor being produced by a separate piece of
the apparatus. The apparatus can then be regarded as two identifiably
distinct pieces of equipment, and therefore represented by a factorizable
element of $\mathbb{H(\mathcal{H})}$. Likewise, it is possible to perform a
single experiment consisting of two pieces of equipment placed at large
spacelike separations, such that each part of the experiment acts on some
aspect of an entangled state. The combined pieces of equipment can in some
circumstances be regarded as separable, and in other circumstances, as an
entangled pair. Certainly, if the equipment is designed to have entangled
outcomes, as recent teleportation experiments are designed to do, then the
two pieces of equipment cannot be regarded as separable, because by our
discussion above, such equipment could only produce separable states.

Theorem $4$ underlines the fact that the operator causal set structure
constrains the state causal set structure, as the following examples show.

\begin{enumerate}
\item[\textbf{Example 3:}]  (Quantum Big Bang) Consider a universe
consisting of a vast number $N=2^{M}$ of qubits, where $M$ is an integer and
assume that the stages paradigm holds. Then each state of the universe $\Psi
_{n}$ is an eigenstate of some test $\hat{O}_{n}$ and the total quantum
register $\mathcal{H}_{[1\ldots N]}$ has dimension $2^{2^{M}}.$

Suppose the state of the universe $\Psi _{n}$ is fully entangled for $n<0$.
Then according to Theorem $4$, it is necessarily the case that for $n<0$, $%
\hat{O}_{n}$ is fully entangled, relative to the skeleton set associated
with the register.

Now suppose that for each time $n$ after zero, the stages dynamics is such
that $\hat{O}_{n}$ doubles the number of its factors, according to the
scheme
\begin{eqnarray}
\hat{O}_{0} &=&\hat{A}^{1\ldots 2^{M}}\in \mathbb{H}\left( \mathcal{H}%
^{1\ldots 2^{M}}\right) ,  \notag \\
\hat{O}_{1} &=&\hat{A}^{1\ldots 2^{M-1}}\otimes \hat{A}^{(2^{M-1}+1)\ldots
2^{M}}\in \mathbb{H}\left( \mathcal{H}^{1\ldots 2^{M-1}\bullet
(2^{M-1}+1)\ldots 2^{M}}\right) ,  \notag \\
\hat{O}_{2} &=&\hat{A}^{1\ldots 2^{M-2}}\otimes \hat{A}^{(2^{M-2}+1)\ldots
2^{M-1}}\otimes \hat{A}^{(2^{M-1}+1)\ldots 3\times 2^{M-2}}\otimes \hat{A}%
^{(3\times 2^{M-2}+1)\ldots 2^{M}}  \notag \\
&&\vdots  \notag \\
\hat{O}_{M} &=&\hat{A}_{1}\otimes \hat{A}_{2}\otimes \ldots \otimes \hat{A}%
_{2^{M}},\;\;\;\hat{A}_{i}\in \mathbb{H}\left( \mathcal{H}_{i}\right) ,
\end{eqnarray}
up to the maximum possible. Then according to Theorem $4$, whatever the
outcome $\Psi _{n}$ is, it must increase its separability at each stage
after exo-time zero until it attains total factorizability at exo-time $M.$
Unless $\hat{O}_{n}$ entangles in any way after time $M$, the fate of this
model universe is a ``heat death'', where the universe ends up consisting of
$2^{M}$ non-interacting, totally isolated qubit sub-universes.

In this scheme, the operators are driven relentlessly to factorize after
time zero, regardless of state outcome. Therefore there is no ``feedback''
from the outcomes. An alternative scheme with such a feedback would be to
have factorization of the operators dependent on whether an outcome was
itself separable or entangled.

\item[\textbf{Example 4:}]  (EPR experiments) Suppose an entangled state $%
\Psi \in \mathcal{H}^{12}$ is prepared via some observable $\hat{O}$. Then $%
\hat{O}$ necessarily has to be entangled, according to Theorem 4.

Suppose now that subsequently, $\hat{O}$ factorizes into two operators, i.e.
$\hat{O}\rightarrow \hat{A}\otimes \hat{B}.$ Then the outcome of this new
test is necessarily separable, i.e. $\Psi \rightarrow \Theta _{A}\otimes
\Theta _{B}$, where $\Theta _{A}$ is an eigenstate of $\hat{A}$ (Alice) and $%
\Theta _{B}$ is an eigenstate of $\hat{B}$ (Bob). From this we see that
quantum entanglement can be ``unravelled'' by ensuring that entangled states
are tested by apparatus consisting of distinct pieces. Conventionally, this
can be arranged by separating such pieces in physical space, but that is
only something that can be considered in the emergent limit where physical
space is a good approximation.

\item[\textbf{Example 5:}]  (Superluminal correlations) Suppose we have a
quantum register consisting of a large number $2N$ of qubits, and suppose
that at time $n$, the state $\Psi _{n}$ of the universe consists of two
factors, $\psi ^{1\ldots N}\in $ $\mathcal{H}^{1\ldots N}$ and $\phi
^{(N+1)\ldots 2N}\in \mathcal{H}^{\left( N+1\right) \ldots 2N}.$ By our
notation, this means that each of these is completely entangled with respect
to its half of the quantum register. Suppose further that $\psi ^{1\ldots N}$
is an eigenstate of factor operator $\hat{A}^{1\ldots N}\in \mathbb{H}\left(
\mathcal{H}^{1\ldots N}\right) $ and that $\phi ^{(N+1)\ldots 2N}$ is an
eigenstate of factor operator $\hat{B}^{\left( N+1\right) \ldots 2N}\in
\mathbb{H}\left( \mathcal{H}^{\left( N+1\right) \ldots 2N}\right) .$

Now suppose that qubit $N+1$ joins qubits $1$ to $N$ to form a factor $\hat{C%
}^{1\ldots \left( N+1\right) }\in \mathbb{H}\left( \mathcal{H}^{1\ldots
\left( N+1\right) }\right) $ in the next test $\hat{O}_{n+1}$ of the
universe and that $\Theta $ is one of the eigenstates of $\hat{O}_{n+1}$.
Then the probability $P(\Theta |\Psi _{n},\hat{O}_{n+1})$ that $\Theta $ is
the next state of the universe cannot factorize in any way, i.e., has only
one factor. Essentially, any potential family structure is totally destroyed
as far as the states are concerned, whereas from the point of view of the
operators, the change appears relatively insignificant. This is the hall
mark of the difference between Einstein locality and superluminal quantum
correlations. Small changes consistent with the principles of relativity in
operators can have consequences which reach across the universe as far as
the states are concerned.
\end{enumerate}

From examples such as these, we are led to believe that the mathematical
differences between $\mathcal{H}$ and $\mathbb{H}(\mathcal{H})$ will provide
important constraints on the natures of the causal sets concerned and
possibly explain why one of them, based on the operators, might satisfy
Einstein locality conditions whilst the other one, based on the states, need
not.

It is our proposal that Einstein causality structure emerges at the point
where tests factorize relative to the basic skeleton set associated with the
fundamental basis for $\mathcal{H}$. This is based on the observation that
classical cellular automata have causal structures quite analogous to light
cones in relativity, provided their rules are local, that is, use
information from neighbouring cells, which are defined in terms of close
family relationships, such as between parents and offspring. \ Locality in
this context is regarded as synonymous with separability, whilst
non-locality is related to entanglement.

In such automata, the causal relationships between neighbouring cells can
propagate signals along what effectively plays the role of light cones. In
our scenario, it would be the pattern of such behaviour in the operators
which effectively generates Einstein locality and which subsequently drives
analogous patterns for the state outcomes, whenever quantum correlation
effects can be ignored.

\

\begin{center}
{\large {\textbf{X. CONCLUDING REMARKS}} }
\end{center}

In this article, our aim of trying to understand the origin of causal set
structure in the universe has rested on two hypotheses: one, that the
universe is a fully self-contained quantum system, and two, that it is
described in terms of a large but finite tensor product of elementary
quantum subregisters, or qubits. We have found that these assumptions lead
in a natural way to a picture of a quantum universe in which factorization
and entanglement of states and observables can provide a basis for a causal
set picture to emerge. Central to our discussion is von Neumann's state
reduction concept, which we believe is not an ugly and ad hoc blemish on the
face of Schr\"{o}dinger mechanics (as the many-worlds and decoherence
theorists would have us believe), but a necessary concept directly relevant
to what happens in the laboratory and the wider universe. It encodes quantum
principles concerning the acquisition of information and underpins for
example the ``no-cloning'' theorem. Given a state $\Psi$, we may attempt to
extract information such as position by a number of position tests. If state
reduction did not occur during each individual outcome, so that $\Psi$ was
invariant to each test of position, then we could proceed to extract
momentum information (say), and end up having more information about the
state than the uncertainty principle permits.

We have outlined a general framework, but it is clear that many details
await further investigation. The programme of emergence, that is, the
explanation of how the universe that we believe we see arises from our
paradigm, is a hard problem which will take a long time to explore in any
detail. It touches upon a major area of physics which has been virtually
unexplored in any detail to date, that is, the description of physics from
within, i.e., endo-physics. This must inevitably be the correct approach to
any investigation which attempts to discuss the universe in a consistent way.

\

\begin{center}
\textbf{ACKNOWLEDGMENTS}
\end{center}

J.E. is grateful to E.P.S.R.C (UK) for a research studentship.

\newpage

\begin{center}
{\large {\textbf{APPENDIX: THE STAGES PARADIGM}} }
\end{center}

In this section we review the stages paradigm \cite
{JAROSZKIEWICZ-01A,JAROSZKIEWICZ-02A} in its general form, which is a
mathematical framework describing the dynamical behaviour of a fully
quantized, self contained and self-referential universe. By this is meant
the extension of the standard principles of quantum mechanics to encompass
the entire universe. In this paradigm the universe behaves as a quantum
automaton, that is, a generalized quantum computation in a vast Hilbert
space composed of many quantum subregisters.

\begin{enumerate}
\item  Spacetime as a manifold does not exist \emph{per se} but is an
emergent concept, as are concepts of metric, dimensionality of space,
reference frames and observers. Pre-emergent time, or ``exo-time'', is
synonymous with the quantum process of change of one stage of the universe
to the next and is discrete, the origin of this temporal discreteness being
quantum state reduction. It follows that successive stages may be labelled
by an integer $n$.

\item  At any given time $n$, the universe is in a unique stage $\Omega
_{n}\equiv \Omega \left( \Psi _{n},\mathbb{I}_{n},\mathcal{R}_{n}\right) $
which has three essential components:

\begin{enumerate}
\item[i)]  $\Psi_{n}$, the current \emph{state of the universe}, is an
element in a Hilbert space $\mathcal{H}$ of enormous dimension $N\gg1.$ The
emergence of classical space and the separability of physical systems in the
universe support the assumption that $\mathcal{H}$ is a tensor product of a
very large number of elementary Hilbert spaces, such as qubits. Any state of
the universe is always a pure state and there are no external observers.

\item[ii)]  $\mathbb{I}_{n}$, the current \emph{information content} is
information over and above that contained in $\Psi _{n},$ such as which test
(see below) $O_{n}$ produced $\Psi _{n}$. $\mathbb{I}_{n}$ is needed for the
dynamics and is classical in that it can be regarded as certain insofar as
the dynamical rules are concerned governing the future evolution of the
universe;

\item[iii)]  $\mathcal{R}_{n}$, the current \emph{rules}, govern the
dynamical development of the universe;
\end{enumerate}

\item  For any given stage $\Omega _{n},$ all other stages such as $\Omega
_{n+1},\Omega _{n-1}$ can only be discussed in terms of \emph{conditional
probabilities, }relative to the condition that the universe is in stage $%
\Omega _{n}.$ This is a mathematical representation of the concept of \emph{%
process time};

\item  The dynamics in the paradigm follows all of the standard principles
of quantum mechanics \cite{PERES:93}, except for the non-existence of
semi-classical observers with free will, and occurs as follows:

i) The current state of the universe (referred to as the \emph{present}) $%
\Psi _{n}$ is the unique outcome (modulo inessential phase) of some unique
test $O_{n},$ represented by a strong element $\hat{O}_{n}$ of $\mathbb{H}%
\left( \mathcal{H}\right) ,$ the set of Hermitian operators on $\mathcal{H}$%
. $\Psi _{n}$ acts as the initial state for the next test, represented by $%
\hat{O}_{n+1},$ which is also a strong element of $\mathbb{H}\left( \mathcal{%
H}\right) $.

ii) As a strong operator, $\hat{O}_{n+1}$ is associated with a unique
preferred basis, $\frak{B}_{n+1}$, which consists of the eigenstates of $%
\hat{O}_{n+1}$. These form a complete orthonormal set and the next state of
the universe $\Psi _{n+1}$ is one of these possible eigenstates.

iii) The factors which determine $\hat{O}_{n+1}$ depend \emph{only} on $%
\Omega _{n}\equiv \Omega \left( \Psi _{n},\mathbb{I}_{n},\mathcal{R}%
_{n}\right) $ and are currently not understood, but do \emph{not} involve
any external observer making a free choice. Given $\hat{O}_{n+1}$, however,
the conditional probability $P\left( \Psi _{n+1}=\Phi |\Psi _{n},\hat{O}%
_{n+1}\right) $ that the next state of the universe $\Psi _{n+1}$ is a
particular eigenstate $\Phi $ of $\hat{O}_{n+1}$ is given by the standard
quantum rule due to Born$,$ i.e.
\begin{equation}
P\left( \Psi _{n+1}=\Phi |\Psi _{n},\hat{O}_{n+1}\right) =|\langle \Phi
|\Psi _{n}\rangle |^{2},  \notag  \label{222}
\end{equation}
assuming the vectors $\Psi _{n}$, $\Phi $ are normalized to unity.

More generally, if we do \emph{not} know what $\hat{O}_{n+1}$ is, we should
use the full stage-stage probability
\begin{equation}
P\left( \Omega _{n+1}^{A\alpha }|\Omega _{n}\right) =|\langle \Phi ^{A\alpha
}|\Psi _{n}\rangle |^{2}P\left( \hat{O}_{n+1}^{A}|\Omega _{n}\right) ,
\notag
\end{equation}
where $P\left( \hat{O}_{n+1}^{A}|\Omega _{n}\right) $ is the conditional
probability that $\hat{O}_{n+1}^{A}$ is selected from $\mathbb{H}\left(
\mathcal{H}\right) $ given $\Omega _{n}$, and $\Phi ^{A\alpha }\in \mathcal{H%
}$ is one of the eigenstates of $\hat{O}_{n+1}^{A}$.

These probabilities are meaningful only from the point of view of
endo-physical observers (macroscopic patterns of factorization of states and
observables) such as physicists who are attempting to understand what the
future of the universe may be like.

\item  Although stages appear to jump in a serial and absolute way, as
labelled by exo-time, an important caveat to this idea involves the concept
of \emph{null test} \cite{JAROSZKIEWICZ-01A}, which allows for a
``multi-fingered'' view of time. Under some circumstances, parts of the
universe generated by long gone stages and represented by certain factors in
$\Psi _{n}$ may remain ``frozen'' for many successive jumps of the universe
and change only during some future test. Such a possibility arises if the
state of the universe is factorizable, which we assume here. When some of
these factors remain unchanged from jump to jump, the result is effectively
one where time regarded in terms of information change appears local. This
internal time is called ``endo-time'', and it is non-integrable, i.e., it is
path-dependent, unlike exo-time. On emergent scales this should provide a
dynamical origin for classical general relativity, including special
relativity as a special case. In mathematical terms, this phenomenon
originates in the fact that a quantum test of a state does not alter that
state if the state is already an eigenstate of the test \cite{DIRAC:58}.
Such a test has no real physical content and leads to no change of
information in any part of the universe;

\item  After each jump $\Psi _{n}\rightarrow \Psi _{n+1},$ the information
content $\mathbb{I}_{n}$ and the rules $\mathcal{R}_{n}$ are updated to $%
\mathbb{I}_{n+1}$ and $\mathcal{R}_{n+1}$ respectively and the whole process
is repeated. According to R. Buccheri \cite{BUCCHERI-02}, how the rules
might change must somehow be encoded into the rules themselves, i.e., they
are also the ``rules of the rules''.
\end{enumerate}


\newpage

\begin{thebibliography}{99}
\bibitem{WHEELER-80}  J. A. Wheeler, in \emph{Quantum Theory and Gravitation}%
, (Academic Press, Inc, New York, 1980), edited by A.R. Marlow.

\bibitem{STUCKEY-99}  M. Stuckey, in \emph{Studies of the Structure of Time:
from Physics to Psycho(Patho)Logy}, edited by R. Buccheri, V. Di Ges\`{u}
and M. Saniga (Kluwer, Dordrecht, 1999), 121.

\bibitem{SORKIN+al-87}  L. Bombelli, J. Lee, D. Meyer, and R.D. Sorkin,
Phys. Rev. Lett \textbf{59}, 521 (1987).

\bibitem{BRIGHTWELL-91}  G. Brightwell and R. Gregory, Phys. Rev. Lett
\textbf{66}, 260 (1991).

\bibitem{SORKIN-99}  D. P. Rideout and R.D. Sorkin, Phys. Rev D \textbf{61},
024002 (2000).

\bibitem{SORKIN+RIDOUT-00}  D.P. Rideout and R.D. Sorkin, Phys. Rev D
\textbf{63}, 104011 (2001).

\bibitem{BRIGHTWELL-02}  G. Brightwell \emph{et al}, arXiv:gr-qc/0202097
(2002), and Proceedings of Alternative Natural Philosophy Association
Meeting, August 16-21, Cambridge, UK (2001)

\bibitem{REGGE-61}  T. Regge, Il Nuovo Cimento \textbf{19}(3), 558 (1961).

\bibitem{PLENIO-98}  M.B. Plenio and P. L. Knight, Rev. Mod. Phys.\textbf{70}%
, 101 (1998)

\bibitem{MARKOPOULOU-00}  F. Markopoulou, Class. Quant. Grav \textbf{17}
2059 (2000).

\bibitem{REQUARDT-99}  M. Requardt, arXiv:gr-qc/9902031 (1999).

\bibitem{HOWSON:72}  A.G. Howson, \emph{A Handbook of Terms Used in Algebra
and Analysis} (Cambridge University Press, 1972).

\bibitem{PRICE:97}  H. Price, \emph{Time's Arrow} (Oxford University Press,
1997).

\bibitem{PERES:93}  A. Peres, \emph{Quantum Theory: Concepts and Methods}
(Kluwer Academic Publishers, 1993).

\bibitem{FINK+LESCHKE-00}  H. Fink and H. Leschke, Found. Phys. Lett \textbf{%
13}, 345 (2000).

\bibitem{BREUER-95}  T. Breuer, Phil. Science \textbf{62}, 197 (1995).

\bibitem{DeWITT+GRAHAM:73}  B. S. DeWitt and N. Graham, \emph{The
Many-Worlds Interpretation of Quantum Mechanics} (Princeton University
Press, 1973).

\bibitem{DEUTSCH:97}  D. Deutsch, \emph{The Fabric of Reality} (The Penguin
Press, 1997).

\bibitem{DEUTSCH-01}  D. Deutsch, arXiv:quant-ph/0104033 (2001).

\bibitem{JAROSZKIEWICZ-01A}  G. Jaroszkiewicz, arXiv:quant-ph/0105013 (2001).

\bibitem{JAROSZKIEWICZ-03A}  J. Eakins and G. Jaroszkiewicz, J. Phys. A:
Math. Gen. \textbf{36}, 517 (2003).

\bibitem{KOCHEN+SPECKER-67}  S. Kochen and E. Specker, J. Math. and Mech.
\textbf{17}, 59 (1967).

\bibitem{SCARANI-00}  V. Scarani, W. Tittel, H. Zbinden and N. Gisin, Phys.
Lett. A \textbf{276}, 1 (2000)

\bibitem{BJORKEN+DRELL:65B}  J. D. Bjorken and S. D. Drell, \emph{%
Relativistic Quantum Fields} (McGraw-Hill Inc, 1965).

\bibitem{GASIOROWICZ:68}  S. Gasiorowicz, \emph{Elementary Particle Physics}
(John Wiley and Sons, 1967).

\bibitem{ZIZZI-01}  P. A. Zizzi, arXiv:gr-qc/0103002 (2001).

\bibitem{JAROSZKIEWICZ-02A}  J. Eakins and G. Jaroszkiewicz,
arXiv:quant-ph/0203020 (2002).

\bibitem{DIRAC:58}  P.A.M. Dirac, \emph{The Principles of Quantum Mechanics}
(Clarendon Press, 1958).

\bibitem{BUCCHERI-02}  R. Buccheri, in \emph{The Nature of Time: Geometry,
Physics and Perception} (Kluwer Dordrecht, 2003), edited by R. Buccheri, M.
Saniga and M. Stuckey.
\end{thebibliography}
\end{document}